\documentclass[letterpaper,11pt]{article}
\pdfoutput=1 

\usepackage{jheppub} 

\usepackage[T1]{fontenc} 
\usepackage{lmodern}
\usepackage[mathscr]{euscript}
\usepackage{graphicx}
\usepackage{ytableau}
\usepackage{amsmath}
\usepackage{amssymb}
\usepackage{hyperref}
\usepackage{xcolor}
\usepackage{color}
\usepackage{dsfont}
\usepackage{graphicx}
\usepackage{tikz}
\usetikzlibrary{shapes.geometric}
 
\usepackage{extarrows} 
\definecolor{navyblue}{rgb}{0.0, 0.0, 0.5}
\definecolor{firebrick}{rgb}{0.7, 0.13, 0.13}
\usetikzlibrary{shapes.misc}

\DeclareFontFamily{U}{mathx}{\hyphenchar\font45}
\DeclareFontShape{U}{mathx}{m}{n}{
      <5> <6> <7> <8> <9> <10>
      <10.95> <12> <14.4> <17.28> <20.74> <24.88>
      mathx10
      }{}
\DeclareSymbolFont{mathx}{U}{mathx}{m}{n}
\DeclareFontSubstitution{U}{mathx}{m}{n}
\DeclareMathSymbol{\bigplus}        {1}{mathx}{"90}
\DeclareMathSymbol{\bigtimes}       {1}{mathx}{"91}

\DeclareFontFamily{OT1}{pzc}{}
\DeclareFontShape{OT1}{pzc}{m}{it}{<-> s * [1.10] pzcmi7t}{}
\DeclareMathAlphabet{\mathpzc}{OT1}{pzc}{m}{it}

\def\beq{\begin{equation}}
\def\eeq{\end{equation}}

\title{\boldmath
{SCFT/VOA correspondence via $\Omega$-deformation}}

\author{Saebyeok Jeong}
\preprint{YITP-SB-19-7}

\affiliation{C.N. Yang Institute for Theoretical Physics, Stony Brook University,\\Stony Brook, NY 11794-3840, USA}

\emailAdd{saebyeok.jeong@gmail.com}

\abstract{We investigate an alternative approach to the correspondence of four-dimensional $\EuScript{N}=2$ superconformal theories and two-dimensional vertex operator algebras, in the framework of the $\Omega$-deformation of supersymmetric gauge theories. The two-dimensional $\Omega$-deformation of the holomorphic-topological theory on the product four-manifold is constructed at the level of supersymmetry variations and the action. The supersymmetric localization is performed to achieve a two-dimensional chiral CFT. The desired vertex operator algebra is recovered as the algebra of local operators of the resulting CFT. We also discuss the identification of the Schur index of the $\EuScript{N}=2$ superconformal theory and the vacuum character of the vertex operator algebra at the level of their path integral representations, using our $\Omega$-deformation point of view on the correspondence.}

\begin{document} 
\maketitle
\flushbottom
\usetikzlibrary{calc,decorations.markings}

\section{Introduction} \label{sec:intro}
Superconformal field theories (SCFTs) exhibit interesting aspects and rich structures due to their large symmetry group. A striking feature revealed in \cite{ref:bllprr} is that any superconformal field theory with an $\mathfrak{su}(1,1 \vert 2)$ superconformal subalgebra which acts as anti-holomorphic M\"{o}bius transformations on a two-dimensional plane possesses a protected sector isomorphic to a two-dimensional vertex operator algebra (VOA).\footnote{We use the terms \textit{vertex operator algebra} and \textit{chiral algebra} interchangeably.} The protected sector is formed as a certain ($\mathcal{Q}+\mathcal{S}$)-cohomology, spanned by twisted-translations of Schur operators with their operator product expansions (OPEs) in the cohomology. 

For Lagrangian four-dimensional $\EuScript{N}=2$ superconformal theories, the procedure of obtaining this chiral algebra can be briefly described as follows. It can be shown that chiral algebras produced by free hypermultiplet and free vector multiplet are those of symplectic bosons (also known as $\beta\gamma$ system) and $bc$ ghosts, respectively. When they are coupled to produce an interacting SCFT, the prescription is first to take the naive tensor product of those two-dimensional chiral algebras with the gauge-invariance constraint and then to pass to the cohomology of the nilpotent BRST operator. Such a procedure led to many conjectural relations in \cite{ref:bllprr} between $\EuScript{N}=2$ superconformal QCDs and $\mathcal{W}$-algebras, which were checked at the level of the equivalence of the superconformal indices and the vacuum characters. For related works, see also \cite{ref:bprvr, ref:brvr, ref:lrs, ref:ll}.

The protected chiral algebra is particularly interesting since it is a non-commutative algebra of local operators in two dimensions, which is not easily expected for theories in higher dimensions. It turns out that the \textit{non-commutative deformation} parameter $\hbar$, which appears in the numerators of the OPEs of chiral algebra, is given by the relative coefficient of the combination $\mathcal{Q}+\mathcal{S}$. Even though this is a direct consequence of OPE computation, it seems that an intuitive understanding of the appearance of the non-commutative deformation parameter is still absent. Therefore, it could be useful to approach the mentioned chiral algebra in an alternative framework where the origin of the non-commutative deformation parameter is well understood. The main goal of the present work is to make such an attempt.

The framework that we are referring to is the $\Omega$-deformation of supersymmetric gauge theories \cite{ref:nek1, ref:nekokoun, ref:nekwit}. It was firstly introduced in \cite{ref:nek1} to regularize the partition function of $\EuScript{N}=2$ gauge theories on the non-compact $\mathbb{C}^2$. Essentially, the $\Omega$-deformation is implemented by modifying the theory as a cohomological field theory with respect to the supersymmetry which squares to an isometry of the underlying manifold. It effectively turns on a potential along the direction orthogonal to the isometry, and thus \textit{localizes} the theory on the fixed points of the isometry. A remarkable discovery made in \cite{ref:neksha} was that the two-dimensional $\Omega$-deformation on $\EuScript{N}=2$ gauge theories can be used to quantize the classical integrable system whose Hamiltonians are given by the $\EuScript{N}=2$ chiral operators. One may regard this quantization at the level of the representations of the non-commutative deformation of the algebra of holomorphic functions on the phase space of the integrable system, where the non-commutative deformation paramter is identified with the $\Omega$-deformation parameter $\varepsilon=\hbar$. A similar feature is also present in other contexts: in three-dimensional $\EuScript{N}=4$ theories, for example, the $\Omega$-deformation on the Rozansky-Witten theory leads to a non-commutative deformation of the Higgs branch chiral ring \cite{ref:yagi,ref:bdg,ref:bbzbdn}.

For which theory should we implement the $\Omega$-deformation to recover the chiral algebra? In \cite{ref:kap}, Kapustin discussed the holomorphic-topological twist of $\EuScript{N}=2$ gauge theories on a product manifold $\EuScript{C} \times \EuScript{C}^\perp$, in which the theory is topological along, say, $\EuScript{C}^\perp$ and holomorphic along $\EuScript{C}$ (see also \cite{ref:aj, ref:nekthe,ref:bln} for earlier works on partially holomorphic and partially topological theories). The cohomology of local operators, therefore, forms a chiral algebra on $\EuScript{C}$, albeit a \textit{commutative} one since local operators can commute with each other by escaping to the direction of $\EuScript{C}^\perp$. Now we can imagine implementing the $\Omega$-deformation with respect to the isometry on $\EuScript{C}^\perp$, effectively creating a potential along the direction of $\EuScript{C}^\perp$. As local operators are now trapped on $\EuScript{C}$ due to the potential, it is natural to expect that we obtain a non-commutative deformation of the chiral algebra. The \textit{height} of the potential would be controlled by none other than the $\Omega$-deformation parameter, and we expect the identification of the non-commutative deformation parameter with the $\Omega$-deformation parameter. We will see that this is indeed the case.

To obtain the two-dimensional chiral algebra, we have to perform supersymmetric localization of the $\Omega$-deformed holomorphic-topological theory to produce a chiral CFT on $\EuScript{C}$. The algebra of local operators of this CFT would provide our desired chiral algebra. It turns out that the localization procedure can be conducted in a very similar manner with \cite{ref:nek}, where the localization of the $\Omega$-deformed two-dimensional Landau-Ginzburg model was discussed. In fact, our localization can be viewed as the gauge theory analogue of \cite{ref:nek} on $\EuScript{C}^\perp$, which was discussed in \cite{ref:cosyagi} in its application of recovering four-dimensional Chern-Simons theory from six-dimensional supersymmetric gauge theory (see also \cite{ref:yagi, ref:ltyz} for the discussion of $B$-models on the compact disk where the localization locus was chosen to be constant maps), occuring at each point of $\EuScript{C}$. The localization locus is given by solutions to certain gradient flow equations (emanating from the critical point of the superpotential as we take $\EuScript{C}^\perp = \mathbb{R}^2$). To obtain the action of the localized theory on $\EuScript{C}$, we have to evaluate the action on this localization locus. This can be accomplished with the help of the equivariant integration, in a similar manner that \cite{ref:nekmar} applies an equivariant integration on $\mathbb{C}^2$ to yield the representations of $\EuScript{N}=2$ chiral operators on the instanton moduli space. For the case at hand, it turns out that there is no non-trivial topological sector of gauge field configurations in the localization locus, so that the further integration on the instanton moduli space would not take place.

The paper is organized as follows. In section \ref{sec:supalg}, we briefly review the Donaldson-Witten twist and the holomorphic-topological twist of Kapustin for four-dimensional $\EuScript{N}=2$ theories. In section \ref{sec:local}, we perform the supersymmetric localization of the $\Omega$-deformed holomorphic-topological theory to obtain the two-dimensional chiral CFT.  In section \ref{sec:ind}, we discuss the identification of $S^3 \times S^1$ partition function of $\EuScript{N}=2$ SCFT and torus partition function of chiral CFT, which lead to the equivalence of the Schur index and the vacuum character. We conclude in section \ref{sec:dis} with discussions.

\paragraph{Note added.} When this work has been completed, we became aware of \cite{ref:ohyagi} which overlaps with the contents of our paper.

\section{Holomorphic-topological twist of $\EuScript{N}=2$ theories} \label{sec:supalg}
Let us consider a $\EuScript{N}=2$ supersymmetric theory on a four-dimensional Euclidean manifold, $X = \EuScript{C} \times \EuScript{C}^\perp$, where $\EuScript{C}$ and $\EuScript{C}^\perp$ are Riemann surfaces. A curved background on $X$ would generically break all the supersymmetries. To preserve some supersymmetries, we need to twist the holonomy group with the R-symmetry group, for which the supercharges with charge $0$ under the twisted holonomy group would remain preserved.

The holonomy group of $X$ is $U(1)_{\EuScript{C}} \times U(1)_{\EuScript{C}^\perp}$ and the R-symmetry group of a $\EuScript{N}=2$ supersymmetric theory is $SU(2)_R \times U(1)_r$. The $\EuScript{N}=2$ superalgebra contains the following supercharges
\begin{align}
\mathcal{Q}^A _\alpha, \quad \widetilde{\mathcal{Q}}^A _{\dot{\alpha}}, \quad A=1,2, \, \alpha=\pm,\, \dot{\alpha} = \dot{\pm},
\end{align}
where $A$ is the $SU(2)_R$ R-symmetry index and $\alpha, \dot{\alpha}$ are un-dotted and dotted spinor indices. We choose the conventions for the generators of the holonomy as 
\begin{align}
\mathcal{M}_{\EuScript{C}} = \mathcal{M}_+ ^{\;\;\; +} + \mathcal{M}^{\dot{+}} _{\;\;\; \dot{+}}, \quad \mathcal{M}_{\EuScript{C}^\perp} = \mathcal{M}_+ ^{\;\;\; +} - \mathcal{M}^{\dot{+}} _{\;\;\; \dot{+}}.
\end{align}
The table \ref{table:charges} shows the supercharges and their quantum numbers. Note that $U(1)_R \subset SU(2)_R$ is the maximal torus.

\begin{table}[h]\centering
\begin{tabular}{l*{8}{c}r}
             & $\mathcal{Q}^1 _+$ & $\mathcal{Q}^1 _-$ & $\mathcal{Q}^2 _+$ & $\mathcal{Q}^2 _-$ & $\widetilde{\mathcal{Q}}^1 _{\dot{+}}$  & $\widetilde{\mathcal{Q}}^1 _{\dot{-}}$ & $\widetilde{\mathcal{Q}}^2 _{\dot{+}}$ & $\widetilde{\mathcal{Q}}^2 _{\dot{-}}$ \\
\hline
$U(1)_{\EuScript{C}}$ & $\frac{1}{2}$ & $-\frac{1}{2}$ & $\frac{1}{2}$ & $-\frac{1}{2}$ & $\frac{1}{2}$ & $-\frac{1}{2}$ & $\frac{1}{2}$ & $-\frac{1}{2}$   \\
$U(1)_{\EuScript{C}^\perp}$            & $\frac{1}{2}$ & $-\frac{1}{2}$ & $\frac{1}{2}$ & $-\frac{1}{2}$ & $-\frac{1}{2}$ & $\frac{1}{2}$ & $-\frac{1}{2}$ & $\frac{1}{2}$  \\
$U(1)_R$           & $\frac{1}{2}$ & $\frac{1}{2}$ & $-\frac{1}{2}$ & $-\frac{1}{2}$ &  $\frac{1}{2}$ & $\frac{1}{2}$ &  $-\frac{1}{2}$  & $-\frac{1}{2}$  \\
$U(1)_r$     & $\frac{1}{2}$ & $\frac{1}{2}$ & $\frac{1}{2}$ & $\frac{1}{2}$ & $-\frac{1}{2}$ & $-\frac{1}{2}$ & $-\frac{1}{2}$ & $-\frac{1}{2}$  \\
\end{tabular} \caption{$\EuScript{N}=2$ supercharges and quantum numbers} \label{table:charges}
\end{table}

\subsection{Donaldson-Witten twist} \label{subsec:dwtw}
Let us first review how the Donaldson-Witten twist comes about. For a curved metric on $\EuScript{C}^\perp$, we twist the holonomy $U(1)_{\EuScript{C}^\perp}$ by taking the diagonal subgroup
\begin{align}
U(1)_{\EuScript{C}^\perp} ' \hookrightarrow U(1)_{\EuScript{C}} \times U(1)_R.
\end{align}
Under the twist, we preserve the $\EuScript{N}=(2,2)$ supersymmetry on $\EuScript{C}$ whose fermionic generators are
\begin{align}
\mathcal{Q}^1 _-,\; \mathcal{Q}^2 _+ ,\; \widetilde{\mathcal{Q}}^1 _{\dot{+}},\; \widetilde{\mathcal{Q}}^2 _{\dot{-}}.
\end{align}
When $\EuScript{C}$ is also curved, we can make a further twist
\begin{align}
U(1)_{\EuScript{C}} ' \hookrightarrow U(1)_{\EuScript{C}} \times U(1)_R
\end{align}
to preserve $\widetilde{\mathcal{Q}}^1 _{\dot{+}}, \; \widetilde{\mathcal{Q}}^2 _{\dot{-}}$. The Donaldson-Witten supercharge is precisely the linear combination of these supercharges,
\begin{align}
\EuScript{Q}_{\text{DW}} = \widetilde{\mathcal{Q}}^1 _{\dot{+}} + \widetilde{\mathcal{Q}}^2 _{\dot{-}}.
\end{align}
Here, $\widetilde{\mathcal{Q}}^1 _{\dot{+}}$ and $\widetilde{\mathcal{Q}}^2 _{\dot{-}}$ are preserved independently but $\EuScript{Q}_{\text{DW}}$ is the one which is preserved for any curved background on $X$, not necessarily a product metric.

To describe the $\Omega$-deformation made upon the twist, let us suppose $\EuScript{C}^\perp = \mathbb{R}^2$ for a moment. One may take a specific combination of supercharges
\begin{align} \label{eq:naive}
\widetilde{\EuScript{Q}} = \widetilde{\mathcal{Q}}^1 _{\dot{+}} +\widetilde{\mathcal{Q}}^2 _{\dot{-}} + \varepsilon (w \mathcal{Q}^1 _{{+}} - \bar{w} \mathcal{Q}^2 _{-} ),
\end{align}
where $w=x^1 + i x^2$ and $\bar{w} = x^1 -ix^2$ are the coordinates on $\EuScript{C}^\perp$. This supercharge squares to the isometry of $\EuScript{C}^\perp$ generated by $V=w\partial_w -\bar{w}\partial_{\bar{w}}$. In general background on $\EuScript{C}^\perp$ the deformed supercharge would not be preserved since the last two supercharges are not preserved as we have seen above. However, one can still construct a deformation of the theory which has a supercharge which squares to the isometry on $\EuScript{C}^\perp$. In practice, we can start from the theory on $\mathbb{R}^4$, write the variations of component fields with respect to the naive supercharge \eqref{eq:naive}, and then seek a way of re-writing them in metric-independent fashion so that deformed supersymmetry variations are consistently defined on arbitrary product manifold $\EuScript{C} \times \EuScript{C}^\perp$. The action of the theory has to be modified correspondingly to ensure the invariance under the deformed supersymmetry.

\begin{table}[h]\centering
\begin{tabular}{l*{8}{c}r}
             & $\mathcal{Q}^1 _+$ & $\mathcal{Q}^1 _-$ & $\mathcal{Q}^2 _+$ & $\mathcal{Q}^2 _-$ & $\widetilde{\mathcal{Q}}^1 _{\dot{+}}$  & $\widetilde{\mathcal{Q}}^1 _{\dot{-}}$ & $\widetilde{\mathcal{Q}}^2 _{\dot{+}}$ & $\widetilde{\mathcal{Q}}^2 _{\dot{-}}$ \\
\hline
$U(1)_{\EuScript{C}} '$ & $0$ & $-1$ & $1$ & 0 & 0 & $-1$ & 1 & 0   \\
$U(1)_{\EuScript{C}^\perp} '$            & 1 & 0 & 0 & $-1$ & 0 & 1 & $-1$ & 0   \\
\end{tabular} \caption{Donaldson-Witten twist} \label{table:dwtwcharge}
\end{table}

\subsection{Holomorphic-topological twist} \label{subsec:holtw}
Now we apply a similar procedure to our main subject: the holomorphic-topological twist of four-dimensional $\EuScript{N}=2$ supersymmetry introduced in \cite{ref:kap}. Let us first breifly review the holomorphic-topological twist. For a curved metric on $\EuScript{C}^\perp$, we twist the holonomy $U(1)_{\EuScript{C}^\perp}$ by taking the diagonal subgroup
\begin{align}
U(1)_{\EuScript{C}^\perp} ' \hookrightarrow U(1)_{\EuScript{C}^\perp} \times U(1)_r.
\end{align}
Under the twist, we preserve the $\EuScript{N}=(0,4)$ supersymmetry on $\EuScript{C}$ whose fermionic generators are
\begin{align}
\mathcal{Q}^A _- , \; \widetilde{\mathcal{Q}}^A _{\dot{-}}, \quad A=1,2.
\end{align}
When $\EuScript{C}$ is also curved, we can make a further twist
\begin{align}
U(1)_{\EuScript{C}} ' \hookrightarrow U(1)_{\EuScript{C}} \times U(1)_R
\end{align}
to preserve $\mathcal{Q}^1 _-, \; \widetilde{\mathcal{Q}}^1 _{\dot{-}}$. The holomorphic-twist supercharge is the following linear combination of supercharges,
\begin{align}
\EuScript{Q} = \mathcal{Q}^1 _- + \widetilde{\mathcal{Q}}^1 _{\dot{-}}.
\end{align}
Note that the translations along $\EuScript{C}^\perp$ and the anti-holomorphic translation along $\EuScript{C}$ are actually $\EuScript{Q}$-exact:
\begin{align}
\begin{split}
&\{ \EuScript{Q} , \mathcal{Q}^2 _+ \} = - \mathcal{P}_{+\dot{-}}, \\
&\{ \EuScript{Q}, \widetilde{\mathcal{Q}}^2 _{\dot{+}} \} = \mathcal{P}_{-\dot{+}} \\
&\{ \EuScript{Q}, \mathcal{Q}^2 _- \} = -\{ \EuScript{Q}, \widetilde{\mathcal{Q}}^2 _{\dot{-}} \} = -\mathcal{P}_{- \dot{-}},
\end{split}
\end{align}
hence it gets the name \textit{holomorphic-topological} twist. Let us suppose $\EuScript{C}^\perp = \mathbb{R}^2$ for a moment. Then we would preserve
\begin{align}
{\EuScript{Q}}_\varepsilon = \mathcal{Q}^1 _- + \widetilde{\mathcal{Q}}^1 _{\dot{-}} + \varepsilon (w \mathcal{Q}^2 _+ + \bar{w} \widetilde{\mathcal{Q}}^2 _{\dot{+}}   ),
\end{align}
which squares to the isometry on $\EuScript{C}^\perp$:
\begin{align}
\EuScript{Q}_\varepsilon ^2 = \varepsilon (w \{ \widetilde{\mathcal{Q}} ^1 _{\dot{-}} , \mathcal{Q}^2 _+ \} + \bar{w} \{ \mathcal{Q}^1 _- , \widetilde{\mathcal{Q}}^2 _{\dot{+}} \} ) = -2\varepsilon (w \mathcal{P}_w -\bar{w} \mathcal{P}_{\bar{w}}).
\end{align}
In general background on $\EuScript{C}^\perp$, the deformed supercharge would not be preserved since the last two supercharges are not preserved as we have seen. However, just as the case of the Donaldson-Witten twist, it is still possible to implement the $\Omega$-deformation of the holomorphic-topological theory by consistently deforming the supersymmetry variations and the action. We will see in the following section how this is actually accomplished.

It is crucial to note that, unlike the Donaldson-Witten case, we make use of the $U(1)_r$ R-symmetry to make a twist with the isometry on $\EuScript{C}^\perp$. Recalling that the deformed supercharge squares to the isometry on $\EuScript{C}^\perp$, we see that the localization with respect to this supercharge would not work if the $U(1)_r$ R-symmetry is anomalous. This is precisely the case when the theory is not superconformal. Thus we restrict our attention to $\EuScript{N}=2$ superconformal theories in relating their $\Omega$-deformation on holomorphic-topological twist with two-dimensional chiral algebras. It is interesting to see that the superconformality is required in a slightly different manner compared to the $(\mathcal{Q}+\mathcal{S})$-cohomology story in \cite{ref:bllprr}, where the superconformal supercharge $\mathcal{S}$ explicitly appears in defining the cohomology of local operators in the chiral algebra.

\begin{table}[h]\centering
\begin{tabular}{l*{8}{c}r}
             & $\mathcal{Q}^1 _+$ & $\mathcal{Q}^1 _-$ & $\mathcal{Q}^2 _+$ & $\mathcal{Q}^2 _-$ & $\widetilde{\mathcal{Q}}^1 _{\dot{+}}$  & $\widetilde{\mathcal{Q}}^1 _{\dot{-}}$ & $\widetilde{\mathcal{Q}}^2 _{\dot{+}}$ & $\widetilde{\mathcal{Q}}^2 _{\dot{-}}$ \\
\hline
$U(1)_{\EuScript{C}} '$ & 1 & 0 & 0 & $-1$ & 1 & 0 & 0 & $-1$   \\
$U(1)_{\EuScript{C}^\perp} '$            & 1 & 0 & 1 & 0 & $-1$ & 0 & $-1$ & 0   \\
\end{tabular} \caption{Holomorphic-topological twist} \label{table:holtwcharge}
\end{table}

\section{Chiral CFT from $\Omega$-deformation and localization} \label{sec:local}
The general analysis of the previous section can be applied to $\EuScript{N}=2$ gauge theories, on which we focus from now on. We perform supersymmetric localization on the $\Omega$-deformed holomorphic-topological theory, to produce a two-dimensional chiral CFT. The desired chiral algebra is obtained as the algebra of local operators of this two-dimensional CFT.

\subsection{Holomorphic-topological twist of $\EuScript{N}=2$ gauge theory}
Let us start from the $\EuScript{N}=2$ vector multiplet. The vector multiplet contains a gauge connection $A$, gaugini $\lambda^A _\alpha$ and $\tilde\lambda^A _{\dot{\alpha}}$, a complex scalar $\phi$, and an auxiliary field $D_{AB}$, where $A=1,2$ is the $SU(2)_R$ R-symmetry index. Following the analysis of the previous section, the holomorphic-topological twist changes the quantum numbers of these component fields as in the table \ref{table:gaugino}. 
\begin{table}[h]\centering
\begin{tabular}{l*{15}{c}r}
             & $\lambda^1 _+$ & $\lambda^1 _-$ & $\lambda^2 _+$ & $\lambda^2 _-$ & $\widetilde{\lambda}^1 _{\dot{+}}$  & $\widetilde{\lambda}^1 _{\dot{-}}$ & $\widetilde{\lambda}^2 _{\dot{+}}$ & $\widetilde{\lambda}^2 _{\dot{-}}$ & & $\phi$ & $\tilde{\phi}$ & & $D^2 _{\;\;2}$ & $D^1 _{\;\;2}$ & $D^2 _{\;\;1}$ \\
\hline
$U(1)_{\EuScript{C}}$ & $\frac{1}{2}$ & $-\frac{1}{2}$ & $\frac{1}{2}$ & $-\frac{1}{2}$ & $\frac{1}{2}$ & $-\frac{1}{2}$ & $\frac{1}{2}$ & $-\frac{1}{2}$ & & 0 & 0  & & 0 &0 &0 \\
$U(1)_{\EuScript{C}^\perp}$            & $\frac{1}{2}$ & $-\frac{1}{2}$ & $\frac{1}{2}$ & $-\frac{1}{2}$ & $-\frac{1}{2}$ & $\frac{1}{2}$ & $-\frac{1}{2}$ & $\frac{1}{2}$ & & 0 & 0 & & 0 &0 &0 \\
$U(1)_R$           & $\frac{1}{2}$ & $\frac{1}{2}$ & $-\frac{1}{2}$ & $-\frac{1}{2}$ &  $\frac{1}{2}$ & $\frac{1}{2}$ &  $-\frac{1}{2}$  & $-\frac{1}{2}$ & & 0 & 0 & & 0 & 1& $-1$ \\
$U(1)_r$     & $-\frac{1}{2}$ & $-\frac{1}{2}$ & $-\frac{1}{2}$ & $-\frac{1}{2}$ & $\frac{1}{2}$ & $\frac{1}{2}$ & $\frac{1}{2}$ & $\frac{1}{2}$ & & $-1$ & 1  &  & 0 & 0& 0 \\
\hline
$U(1)' _{\EuScript{C}}$        & 1 & 0 & 0 & $-1$ & 1 & 0 & 0 & $-1$ & & 0 & 0 & & 0 & 1 & $-1$\\
$U(1)' _{\EuScript{C}^\perp}$       & 0 & $-1$ & 0 & $-1$ & $0$ & 1 & $0$ & 1 & & $-1$ & 1 & & 0 & 0 & 0 \\
\end{tabular} \caption{$\EuScript{N}=2$ vector multiplet; gaugini, scalars, and auxiliary field} \label{table:gaugino}
\end{table}

Correspondingly, we change the notation for the component fields by their representations under the Lorentz group after the twist,
\begin{align} \label{eq:vecdef}
\begin{split}
&\lambda^1 _+ = \lambda_z , \quad \lambda^1 _- = \lambda_{\bar{w}}, \quad \lambda^2 _+ =\lambda, \quad \lambda^2 _- = \lambda_{\bar{z}\bar{w}} , \\
&\tilde\lambda ^1 _{\dot{+}} = \tilde\lambda_z , \quad \tilde\lambda^1 _{\dot{-}} = \tilde\lambda_w, \quad \tilde\lambda^2 _{\dot{+}} = \tilde\lambda, \quad \tilde\lambda^2 _{\dot{-}} = \tilde\lambda_{\bar{z}w} \\
& \phi = \phi_{\bar{w}} ,\quad \tilde\phi =\tilde\phi_w, \quad D^2 _{\;\;2} = D, \quad D^1 _{\;\;2} = D_z, \quad D^2 _{\;\;1} = D_{\bar{z}}.
\end{split}
\end{align}

The $\EuScript{N}=2$ supersymmetry variations can be written as
\begin{align}
\begin{split}
&\delta A_\mu = i \zeta^A \sigma_\mu \tilde\lambda _A -i \tilde\zeta ^A \tilde\sigma_\mu \lambda_A \\
&\delta\phi = -i \zeta ^A \lambda_A \\
&\delta\tilde\phi = i \tilde\zeta ^A \tilde\lambda_A \\
&\delta\lambda_A = \frac{1}{2} F_{\mu\nu} \sigma^{\mu\nu} \zeta_A +2 D_\mu \phi \sigma ^\mu \tilde\zeta_A + \phi \sigma^\mu D_\mu \tilde\zeta_A +2i \zeta_A [\phi,\tilde\phi]+D_{AB} \zeta^B \\
&\delta\tilde\lambda _A = \frac{1}{2}F_{\mu\nu} \tilde\sigma^{\mu\nu} \tilde\zeta _A + 2D_\mu \tilde\phi \tilde\sigma^\mu \zeta_A +\tilde\phi \tilde\sigma^\mu D_\mu \zeta_A -2i \tilde\zeta_A [\phi, \tilde\phi] +D_{AB} \tilde\zeta^B \\
&\delta D_{AB} = -i \tilde\zeta_A \tilde\sigma ^\mu D_\mu \lambda_B +i \zeta_A \sigma ^\mu D_\mu \tilde\lambda_B -2[\phi, \tilde\zeta_A \tilde\lambda_B] +2[\tilde\phi,\zeta_A\lambda_B ] + (A \leftrightarrow B),
\end{split} 
\end{align}
with the fermionic parameters $\zeta^A$ and $\tilde\zeta^A$. In a general metric background, the supersymmetry would be preserved only if $\zeta^A$ and $\tilde{\zeta} ^A$ are Killing spinors. Let us first place the theory on the flat $\mathbb{R}^4$. Since the holomorphic-topological supercharge is $\EuScript{Q} = \mathcal{Q}^1 _- + \widetilde{\mathcal{Q}}^1 _{\dot{-}}$, it is straightforward to write out the variations of the component fields with respect to the holomorphic-topological supercharge, using the notation \eqref{eq:vecdef}, as
\begin{align} \label{eq:susyvec}
\begin{split}
&\EuScript{Q} A_z = \tilde{\lambda}_z -\lambda_z , \quad \EuScript{Q} A_{\bar{z}} = 0, \quad \EuScript{Q} A_w =  \tilde{\lambda}_w , \quad \EuScript{Q} A_{\bar{w}} = -\lambda_{\bar{w}}, \\
&\EuScript{Q} \phi_{\bar{w}} = i \lambda_{\bar{w}} , \quad \EuScript{Q} \tilde{\phi}_w = i \tilde{\lambda}_w, \\
& \EuScript{Q}\lambda_z = D_z, \quad \EuScript{Q}\lambda _{\bar{w}} =0, \quad \EuScript{Q} \lambda _{\bar{z}\bar{w}} =-4F_{\bar{z}\bar{w}} +4i D_{\bar{z}}\phi_{\bar{w}}, \\
&\EuScript{Q}\lambda = 2F_{z\bar{z}} +2F_{w \bar{w}} -4iD_w \phi_{\bar{w}} +2i[\phi_{\bar{w}},\tilde{\phi}_w ] +D, \\
&\EuScript{Q}\tilde{\lambda}_z = D_z , \quad \EuScript{Q}\tilde{\lambda} _w =0, \quad \EuScript{Q}\tilde\lambda _{\bar{z}w} =-4F_{\bar{z}w} -4i D_{\bar{z}} \tilde{\phi}_w, \\
&\EuScript{Q} \tilde{\lambda} = 2F_{z\bar{z}} -2F_{w\bar{w}} +4i D_{\bar{w}} \tilde\phi_w -2i[\phi_{\bar{w}},\tilde\phi_w] +D, \\
&\EuScript{Q} D_z =0, \quad \EuScript{Q}D_{\bar{z}} = 4D_{\bar{z}} (\lambda-\tilde\lambda) +4D_w \lambda_{\bar{z}\bar{w}} -4D_{\bar{w}} \tilde\lambda_{\bar{z}w} +4[\phi_{\bar{w}},\tilde\lambda_{\bar{z}w}] +4[\tilde\phi_w ,\lambda_{\bar{z}\bar{w}}], \\
&\EuScript{Q} D = 2 D_{\bar{z}} (\tilde\lambda_z -\lambda_z) -2D_w\lambda_{\bar{w}} +2D_{\bar{w}} \tilde\lambda_w -2[\phi_{\bar{w}},\tilde\lambda_w] -2[\tilde\phi_w ,\lambda_{\bar{w}}].
\end{split} 
\end{align}
Now we turn to the action for the vector multiplet. It is given by
\begin{align}
\begin{split}
S_{\text{top}} &= -\frac{i \vartheta}{8\pi^2} \int \text{Tr}\, F\wedge F \\
S_{\text{vec}} &= \frac{1}{g^2} \int d^4 x \; \text{Tr}\, \left[ \frac{1}{2}F_{\mu\nu} F^{\mu\nu} -\frac{1}{2}D^{AB} D_{AB} -4D_\mu \tilde\phi D^\mu \phi +4[\phi,\tilde\phi]^2 \right. \\
& \quad\quad\quad \quad\quad\quad\quad\quad  \left. -2 i \lambda^A\sigma^\mu D_\mu \tilde\lambda_A -2\lambda^A [\tilde\phi,\lambda_A]+2 \tilde\lambda^A[\phi,\tilde\lambda_A]\right],
\end{split} 
\end{align}
where $g$ is the gauge coupling. As we will see momentarily, the topological term does not affect the theory and there would be no dependence on $\vartheta$. Thus we drop the topological term from now on. Then a computation shows that the rest of the action turns out to be $\EuScript{Q}$-exact:
\begin{align} \label{eq:vecacta}
\begin{split}
S_{\text{vec}} &= \EuScript{Q} \left[ \frac{1}{g^2} \int d^4 x \;  \text{Tr}\, \left[ 2\lambda_{\bar{z}\bar{w}} (-F_{zw} +i D_z\tilde\phi_w) -2\tilde\lambda_{\bar{z}w} (F_{z\bar{w}}+i D_z\phi_{\bar{w}}) +\frac{1}{2} (\lambda_z +\tilde\lambda_z) D_{\bar{z}} \right. \right. \\
& \quad\quad\quad\quad\quad\quad\quad\quad\quad +(\lambda+\tilde\lambda) \left(-F_{z\bar{z}} +\frac{1}{2} D -iD_{\bar{w}} \tilde\phi_w +i D_w \phi_{\bar{w}} \right) \\
& \quad\quad\quad\quad\quad\quad\quad\quad\quad \left.\left.+(\lambda-\tilde\lambda)\left( -F_{w\bar{w}} -iD_{\bar{w}} \tilde\phi_w -iD_w \phi_{\bar{w}} -i[\phi_{\bar{w}},\tilde\phi_w] \right) \right] \right].
\end{split}
\end{align}
To ensure the positive-definiteness of the action, we impose the reality properties to the bosonic fields,
\begin{align}
\bar{A}_\mu = A_\mu, \quad \bar{ \phi}= - \tilde\phi , \quad \bar{D}_{AB} = - D^{AB},
\end{align}
while requiring the symplectic-Majorana conditions to the gaugini,
\begin{align}
\overline{(\lambda_{A\alpha})} = \epsilon^{AB} \epsilon^{\alpha\beta} \lambda_{B\beta}, \quad \overline{(\tilde\lambda_{A\dot\alpha})} = \epsilon^{AB} \epsilon^{\dot\alpha \dot\beta} \tilde\lambda_{B\dot\beta}.
\end{align}

As mentioned in the previous section, the holomorphic-topological supercharge $\EuScript{Q}=\mathcal{Q}^1 _- + \widetilde{\mathcal{Q}}^1 _{\dot{-}}$ is in fact preserved in any product metric background as long as we make the proper twist of the isometry with the R-symmetry group. Hence we would like to write the supersymmetry variation rules to make sense in a general metric background. This requires a bunch of re-definition of component fields,
\begin{align} \label{eq:redefvec}
\begin{split}
&\phi = \tilde\phi_w dw - \phi_{\bar{w}} d\bar{w}, \quad \mathcal{A} = A+i \phi, \quad \bar{\mathcal{A}} = A-i \phi, \quad \lambda = 2\tilde\lambda_w dw -2 \lambda_{\bar{w}} d\bar{w} \\
&\mu_z = \frac{\lambda_z +\tilde\lambda_z}{2} dw \wedge d\bar{w} ,\quad \mathsf{D}_z = D_z dw \wedge d\bar{w}, \quad \alpha = \frac{\lambda+\tilde\lambda}{2}, \quad \nu= \frac{\lambda-\tilde\lambda}{4} dw \wedge d\bar{w} \\
&\theta_z = \tilde\lambda_z - \lambda_z , \quad {\rho}_{\bar{z}} = \frac{\tilde\lambda_{\bar{z}w} dw + \lambda_{\bar{z}\bar{w}} d\bar{w}}{4}, \quad \mathsf{D}_{\bar{z}} = \frac{1}{16} D_{\bar{z}} dw \wedge d\bar{w} \\
&\mathsf{D} = D+2F_{z\bar{z}} -2iD_w \phi_{\bar{w}} +2i D_{\bar{w}} \tilde\phi_w.\\
\end{split}
\end{align}
For convenience, let us also denote the curvature of the \textit{complexified} connection $\mathcal{A}$, $\bar{\mathcal{A}}$ by
\begin{align}
\begin{split}
&\mathcal{F} = \partial_w \mathcal{A}_{\bar{w}} - \partial_{\bar{w}} \mathcal{A}_w - i [\mathcal{A}_w , \mathcal{A}_{\bar{w}}], \quad\bar{\mathcal{F} }= \partial_w \bar{\mathcal{A}}_{\bar{w}} - \partial_{\bar{w}} \bar{\mathcal{A}}_w - i [\bar{\mathcal{A}}_w , \bar{\mathcal{A}}_{\bar{w}}], \\
& \mathcal{F}_{w\bar{z}} = \partial_w A_{\bar{z}} - \partial_{\bar{z}} \mathcal{A}_w - i [\mathcal{A}_w, A_{\bar{z}}], \quad \mathcal{F}_{\bar{w}\bar{z}} = \partial_{\bar{w}} A_{\bar{z}} - \partial_{\bar{z}} \mathcal{A}_{\bar{w}} - i [\mathcal{A}_{\bar{w}} , A_{\bar{z}}], \\
& \bar{\mathcal{F}}_{wz} = \partial_w A_z - \partial_z \mathcal{A}_z - i[\mathcal{A}_w , A_z],  \quad \bar{\mathcal{F}}_{\bar{w}z} = \partial_{\bar{w}} A_z - \partial_z \mathcal{A}_{\bar{w}} -i[\mathcal{A}_{\bar{w}} , A_z], \\
& \mathcal{F}_{\bar{z}} = \mathcal{F}_{w \bar{z}} dw + \mathcal{F}_{\bar{w}\bar{z}} d\bar{w},\quad \bar{\mathcal{F}}_z = \bar{\mathcal{F}}_{wz} dw + \bar{\mathcal{F}}_{\bar{w}z} d\bar{w} .
\end{split}
\end{align}
Then the supersymmetry variations are significantly simplified in terms of these new fields,
\begin{align} \label{eq:susyvecundef}
\begin{split}
&\EuScript{Q} \mathcal{A} =0, \quad \EuScript{Q} \bar{\mathcal{A}} = \lambda , \\
& \EuScript{Q} \lambda = 0, \quad \EuScript{Q} \nu = \mathcal{F}, \\
& \EuScript{Q} \alpha = \mathsf{D} , \quad \EuScript{Q} \mathsf{D} = 0, \\
& \EuScript{Q} A_{\bar{z}} = 0 ,\quad \EuScript{Q} A_z = \theta_z, \\
& \EuScript{Q} \rho_{\bar{z}} = \mathcal{F}_{\bar{z}}  , \quad \EuScript{Q} \theta_z = 0, \\
& \EuScript{Q} \mathsf{D}_{\bar{z}} = \mathcal{D}_{\EuScript{C}^\perp} \rho_{\bar{z}} +D_{\bar{z}} \nu, \quad \EuScript{Q} \mu_z = \mathsf{D}_z, \quad \EuScript{Q} \mathsf{D}_z = 0,
\end{split}
\end{align}
where we have used the new covariant derivative $\mathcal{D_{\EuScript{C}^\perp}}=d_{\EuScript{C}^\perp} - i \mathcal{A}$ (we also denote $\bar{\mathcal{D}}_{\EuScript{C}^\perp}= d_{\EuScript{C}^\perp} - i \bar{\mathcal{A}}$). The action \eqref{eq:vecacta} for the vector multiplet can be also written in these fields as
\begin{align} \label{eq:actvec}
\begin{split}
S_{\text{vec}}&= \EuScript{Q} \, \Bigg\{ \frac{1}{g^2} \int_{\EuScript{C}} d^2 z \int_{\EuScript{C}^\perp} \text{Tr}\, \left[ -   \bar{\mathcal{F}} \star_{\EuScript{C}^\perp} \nu - \alpha \left( \star_{\EuScript{C}^\perp} \mathsf{D} -2i  D_{\EuScript{C}^\perp} \star_{\EuScript{C}^\perp} \phi -4 \star_{\EuScript{C}^\perp} F_{z\bar{z}} \right) \right. \\
&  \quad\quad\quad\quad\quad\quad\quad\quad\quad\quad \left. +4 \bar{\mathcal{F}}_z \wedge \star_{\EuScript{C}^\perp} \rho_{\bar{z}} + 4 \mu_z \star_{\EuScript{C}^\perp} \mathsf{D}_{\bar{z}} \right] \Bigg\}.
\end{split}
\end{align}

To make a $\EuScript{N}=2$ gauge theory superconformal, we in general need to couple hypermultiplets to the vector multiplet. Let us consider $r$ hypermultiplets which consist of scalars $q_{A I}$, fermions $\psi_{I}, \tilde{\psi}_I$, and auxiliary fields $F_{\check{A}I}$, where $I=1, \cdots, 2r$ is the $Sp(r)$ flavor index. The auxiliary $SU(2)$ $\check {A}=1,2$ is introduced to achieve an off-shell description of the hypermultiplet. We will only use $Sp(1)^r \subset Sp(r)$ subgroup of the flavor symmetry, so let us restrict a single free hypermultiplet $(r=1)$ for a moment. 

Recall that $U(1)_{\EuScript{C}}$ is twisted with the maximal torus of the $SU(2)_R$ R-symmetry group, $U(1)_R \subset SU(2)_R$. For the hypermultiplet, we will take a further twist with the maximal torus of the flavor symmetry:
\begin{align}
U(1)'_{\EuScript{C}} \hookrightarrow U(1)_{\EuScript{C}} \times U(1)_R \times U(1)_{F,\check{F}},
\end{align}
where $U(1)_{F,\check{F}}$ is the maximal torus of the $SU(2)$ flavor group or the $SU(2)$ auxiliary group. This is not really necessary but it will fix the spins of the resulting two-dimenisonal symplectic bosons to be integers. One can always undo this further twist. The tables \ref{table:hyperscal} and \ref{table:hyperfer} show the quantum numbers of the component fields in the hypermultiplet under the twist. 
\begin{table}[h]\centering
\begin{tabular}{l*{4}{c}r}
             & $q_{11}$ & $q_{12}$ & $q_{21}$ & $q_{22}$  \\
\hline
$U(1)_{\EuScript{C}}$ & $0$ & $0$ & $0$ & $0$    \\
$U(1)_{\EuScript{C}^\perp}$            & $0$ & $0$ & $0$ & $0$ \\
$U(1)_R$           & $-\frac{1}{2}$ & $-\frac{1}{2}$ & $\frac{1}{2}$ & $\frac{1}{2}$   \\
$U(1)_r$     & $0$ & $0$ & $0$ & $0$   \\
$U(1)_F$    & $\frac{1}{2}$ & $- \frac{1}{2}$ & $\frac{1}{2}$ & $-\frac{1}{2}$  \\
\hline
$U(1)'_{\EuScript{C}}$ & $0$ & $-1$ & $1$ & $0$    \\
$U(1)'_{\EuScript{C}^\perp}$ & $0$ & $0$ & $0$ & $0$    \\
\end{tabular} \caption{$\EuScript{N}=2$ hypermultiplet, scalars} \label{table:hyperscal}
\end{table}

\begin{table}[h]\centering
\begin{tabular}{l*{8}{c}r}
             & $\psi_{+ 1}$ & $\psi_{- 1}$ & $\psi_{+ 2}$ & $\psi_{- 2}$ & $\tilde{\psi}_{\dot{+}1}$  & $\tilde{\psi}_{\dot{-}1}$ & $\tilde{\psi}_{\dot{+}2}$ & $\tilde{\psi}_{\dot{-}2}$ \\
\hline
$U(1)_{\EuScript{C}}$ & $\frac{1}{2}$ & $-\frac{1}{2}$ & $\frac{1}{2}$ & $-\frac{1}{2}$ & $\frac{1}{2}$ & $-\frac{1}{2}$ & $\frac{1}{2}$ & $-\frac{1}{2}$   \\
$U(1)_{\EuScript{C}^\perp}$            & $\frac{1}{2}$ & $-\frac{1}{2}$ & $\frac{1}{2}$ & $-\frac{1}{2}$ & $-\frac{1}{2}$ & $\frac{1}{2}$ & $-\frac{1}{2}$ & $\frac{1}{2}$   \\
$U(1)_R$           & $0$ & $0$ & $0$ & $0$ &  $0$ & $0$ &  $0$  & $0$  \\
$U(1)_r$     & $\frac{1}{2}$ & $\frac{1}{2}$ & $\frac{1}{2}$ & $\frac{1}{2}$ & $-\frac{1}{2}$ & $-\frac{1}{2}$ & $-\frac{1}{2}$ & $-\frac{1}{2}$ \\
$U(1)_F$    & $\frac{1}{2}$ & $\frac{1}{2}$ & $-\frac{1}{2}$ & $-\frac{1}{2}$ & $\frac{1}{2}$ & $\frac{1}{2}$ & $-\frac{1}{2}$ & $-\frac{1}{2}$  \\
\hline
$U(1)' _{\EuScript{C}}$        & 1 & 0 & 0 & $-1$ & 1 & 0 & 0 & $-1$  \\
$U(1)' _{\EuScript{C}^\perp}$       & 1 & 0 & 1 & 0 & $-1$ & 0 & $-1$ & 0 \\
\end{tabular} \caption{$\EuScript{N}=2$ hypermultiplet, fermions} \label{table:hyperfer}
\end{table}

We define correspondingly
\begin{align}
\begin{split}
&q_z \equiv q_{21}, \quad q_{\bar{z}} \equiv -q_{12} ,  \quad \tilde{q} \equiv q_{22}, \quad \tilde{q}^\dagger = q_{11}, \\
&\psi_{zw} \equiv \psi_{+1} ,\quad \psi_{z \bar{z}} \equiv \psi_{-1}, \quad \tilde{\psi}_{z \bar{w}} \equiv \tilde{\psi}_{\dot{+}1} ,\quad \tilde{\psi}_{z \bar{z}} \equiv \tilde{\psi}_{\dot{-}1} \\
&\psi_w \equiv \psi_{+2}, \quad \psi_{\bar{z}} \equiv \psi_{-2}, \quad \tilde{\psi}_{\bar{w}} \equiv \tilde{\psi}_{\dot{+}2}, \quad \tilde{\psi}_{\bar{z}} \equiv \tilde{\psi}_{\dot{-}2}, \\
& F_z \equiv F_{21}, \quad F_{\bar{z}} \equiv F_{12}, \quad \tilde{F} \equiv F_{22},\quad \tilde{F}^\dagger = - F_{11}.
\end{split}
\end{align}
Let us take the hypermultiplet to be valued in a unitary representation $R$ of the gauge group ($\bar{R}$ denotes the complex conjugate representation which is isomorphic to the dual representation). We take the convention such that the component fields $q_z$, $\tilde{q}^\dagger$, $\psi_{zw}$, $\psi_{z\bar{z}}$, $\tilde\psi_{z\bar{w}}$, $\tilde\psi_{z\bar{z}}$, $F_z$, $\tilde{F}^\dagger$ are valued in $R$ while $q_{\bar{z}}$, $\tilde{q}$, $\psi_w$, $\psi_{\bar{z}}$, $\tilde{\psi}_{\bar{w}}$, $\tilde\psi_{\bar{z}}$, $F_{\bar{z}}$, $\tilde{F}$ are valued in $\bar{R}$. The $\EuScript{N}=2$ supersymmetry variations are given by
\begin{align}
\begin{split}
&\delta q_{AI} = -i \zeta_A \psi_I + i \tilde{\zeta}_A \tilde{\psi}_I \\
&\delta \psi_I = - 2 \sigma^\mu \tilde{\zeta}^A D_\mu q_{A I} +4i \zeta^A (\tilde{\phi} \cdot q_A)_{I} -\sigma^\mu D_\mu \tilde\zeta^A q_{AI} -2\check\zeta^{\check{A}} F_{\check{A}I} \\
&\delta \tilde{\psi}_I = -2 \tilde{\sigma} ^\mu \zeta ^A D_\mu q_{AI} +4i \tilde{\zeta}^A (\phi \cdot q_A)_{I} -\tilde\sigma ^\mu D_\mu \zeta^A q_{AI} -2\tilde{\check\zeta}^{\check{A}} F_{\check{A}I} \\
&\delta F_{\check{A}I} = i\check{\zeta}_{\check{A}} \sigma^\mu D_\mu \tilde\psi_I ^{\dot\alpha} - i \tilde{\check\zeta}_{\check{A}} \tilde\sigma^\mu D_\mu \psi_I -2 ( \phi \cdot \check\zeta_{\check{A}} \psi)_I -2(\check\zeta_{\check{A}} \lambda_B \cdot q^B )_I +2 (\tilde\phi \cdot \tilde{\check\zeta}_{\check{A}} \tilde\psi )_I + 2 (\tilde{\check\zeta}_{\check{A}} \tilde\lambda_B \cdot q^B )_I,
\end{split} 
\end{align}
where the fermionic parameters $\check\zeta_A$, $\tilde{\check\zeta}_A$ should satisfy the constraints
\begin{align} \label{const}
\zeta_A \check\zeta_{\check{B}} - \tilde\zeta_A \tilde{\check\zeta}_{\check{B}} =0, \quad \zeta^A \zeta_A + \tilde{\check\zeta}^{\check{A}} \tilde{\check\zeta}_{\check{A}} =0, \quad \tilde\zeta^A \tilde\zeta_A + \check\zeta^{\check{A}} \check\zeta_{\check{A}} =0, \quad \zeta^A \sigma^\mu \tilde\zeta_A + \check\zeta^{\check{A}} \sigma^\mu \tilde{\check\zeta}_{\check{A}} =0,
\end{align}
to ensure the off-shell invariance of the supersymmetry. Since the holomorphic-topological supercharge is $\EuScript{Q} = \mathcal{Q}^1 _- + \widetilde{\mathcal{Q}}^1 _{\dot{-}}$, we have to find the solutions for $\check\zeta_A$ and $\tilde{\check\zeta}_A$ for $\zeta_1 ^- = 1$ and $\tilde\zeta_1 ^{\dot{-}} =1$. It is not hard to find that $\check\zeta_2 ^+ =1 $ and $\tilde{\check\zeta}_{2\dot{-}} = -1$ satisfy the equations \eqref{const}. Now it is straightforward to write out all the variations of component fields under the action of the holomorphic-topological supercharge:
\begin{align}
\begin{split}
&\EuScript{Q} q_z =0 , \quad \EuScript{Q}\tilde{q} =0, \quad \EuScript{Q} q_{\bar{z}} = i\psi_{\bar{z}} +i \tilde\psi_{\bar{z}}, \quad \EuScript{Q} \tilde{q}^\dagger = -\frac{i}{2} g^{z\bar{z}} (\psi_{z\bar{z}} +\tilde\psi_{z\bar{z}}) \\
&\EuScript{Q}\psi_{zw} = 4i \mathcal{D}_w q_z , \quad \EuScript{Q} \psi_{z\bar{z}} = -4i D_{\bar{z}} q_z +2\tilde{F}^\dagger \\
&\EuScript{Q} \psi_w = 4i \mathcal{D}_w \tilde{q} , \quad \EuScript{Q} \psi_{\bar{z}}= -4i D_{\bar{z}} \tilde{q} -2 F_{\bar{z}} \\
& \EuScript{Q}\tilde\psi_{z\bar{w}} = -4i \mathcal{D}_{\bar{w}} q_z  , \quad \EuScript{Q}\tilde\psi_{z\bar{z}} = 4i D_{\bar{z}} q_z -2 \tilde{F}^\dagger \\
& \EuScript{Q}\tilde\psi_{\bar{w}} = -4i \mathcal{D}_{\bar{w}} \tilde{q}  , \quad \EuScript{Q} \tilde\psi _{\bar{z}} = 4i D_{\bar{z}} \tilde{q} +2F_{\bar{z}} \\
& \EuScript{Q} F_{\bar{z}} = 0 , \quad \EuScript{Q} \tilde{F}^\dagger =0 \\
& \EuScript{Q} F_z = 2\mathcal{D}_{\bar{w}} \psi_{zw}+2\mathcal{D}_w \tilde\psi_{z\bar{w}} -2D_z (\psi_{z\bar{z}} +\tilde\psi_{z\bar{z}})  -2 (\tilde\lambda_z - \lambda_z) \cdot \tilde{q}^\dagger + 2(\lambda -\tilde{\lambda}) \cdot q_z \\
& \EuScript{Q} \tilde{F} =2\mathcal{D}_{\bar{w}} \psi_w +2\mathcal{D}_w \tilde\psi_{\bar{w}} -2D_z (\psi_{\bar{z}} +\tilde\psi_{\bar{z}} )  +2(\tilde\lambda_z -\lambda_z) \cdot q_{\bar{z}} +2(\lambda-\tilde\lambda ) \cdot \tilde{q}.
\end{split}
\end{align}
Finally the action for the hypermultiplet is given by
\begin{align}
\begin{split}
S_{\text{hyp}} =& \frac{1}{g^2} \int d^4 x \left[ \frac{1}{2} D_\mu q^A D^\mu q_A -q^A \{ \phi,\tilde\phi \} q_A +\frac{i}{2} q^A D_{AB} q^B - \frac{i}{2} \tilde\psi \tilde\sigma ^\mu D_\mu \psi - \frac{1}{2} F^{\check{A}} F_{\check{A}} \right. \\
& \quad\quad\quad\quad\quad\quad\quad\quad\quad \left. -\frac{1}{2}\psi \phi \psi +\frac{1}{2} \tilde\psi \tilde\phi \tilde\psi -q^A \lambda_A \psi + \tilde\psi \tilde\lambda_A q^A \right].
\end{split}
\end{align}
To ensure the positive-definiteness of the action, we impose the following reality properties for the scalars,
\begin{align}
\overline{(q_{AI})} = \Omega^{IJ} \epsilon^{AB} q_{BJ}, \quad \overline{(F_{\check{A}I})} = -\Omega^{IJ} \epsilon^{\check{A}\check{B}} F_{\check{B}J},
\end{align}
while requiring the fermions to be $\Omega$-symplectic Majorana
\begin{align}
\overline{(\psi_{{\alpha} I})} = \epsilon^{\alpha\beta} \Omega^{IJ} \psi_{\beta J}, \quad \overline{(\tilde{\psi}_{\dot{\alpha} I})} = \epsilon^{\dot{\alpha}\dot{\beta}} \Omega^{IJ} \tilde{\psi}_{\dot{\beta} J},
\end{align}
where $\Omega^{IJ}$ is the real antisymmetric $Sp(r)$-invariant tensor satisfying
\begin{align}
(\Omega^{IJ})^* = -\Omega_{IJ} , \quad \Omega^{IJ} \Omega_{JK} = \delta^I _K.
\end{align}

Repeating the argument made for the vector multiplet, the holomorphic-topological supercharge is preserved for any product manifold after the twist. Hence we would like to write the supersymmetry variations in metric-independent fashion. This is achieved by making a bunch of re-definition of fields,
\begin{align}
\begin{split}
&\sigma \equiv \frac{1}{4i}(\psi_w dw - \tilde{\psi}_{\bar{w}} d\bar{w}), \quad \xi_z \equiv \frac{1}{4i} (\psi_{zw} dw - \tilde{\psi}_{z \bar{w}} d \bar{w}), \quad \gamma \equiv -\frac{i}{2} g^{z\bar{z}} (\psi_{z\bar{z}} +\tilde\psi_{z\bar{z}}), \\
&\chi \equiv -\frac{ig^{z \bar{z}}(\psi_{z\bar{z}} -\tilde{\psi}_{z\bar{z}})}{4} dw \wedge d\bar{w}, \quad \eta_{\bar{z}} \equiv \frac{i(\psi_{\bar{z}} -\tilde\psi_{\bar{z}})}{2} dw \wedge d\bar{w} ,\quad \zeta_{\bar{z}} \equiv i (\psi_{\bar{z}} +\tilde\psi_{\bar{z}}),  \\
& h_z \equiv \frac{i}{8} (F_z +2i D_z \tilde{q}^\dagger) dw \wedge d\bar{w}, \quad h \equiv \frac{i}{8} (\tilde{F} -i g^{z \bar{z}} D_z q_{\bar{z}} ) dw \wedge d\bar{w}, \\
& h ^\dagger \equiv -2i (\tilde{F}^\dagger -ig^{z\bar{z}} D_{\bar{z}} q_z) dw\wedge d\bar{w} , \quad h_{\bar{z}} \equiv - 2i (F_{\bar{z}} +2i D_{\bar{z}} \tilde{q} ) dw\wedge d\bar{w} .
\end{split} 
\end{align}
In terms of these new fields, the holomorphic-topological supercharge is represented in a simple manner,
\begin{align} \label{eq:susyhypundef}
\begin{split}
& \EuScript{Q} q_z =0, \quad \EuScript{Q} \xi_z =\mathcal{D}_{\EuScript{C}^\perp} q_z, \quad \EuScript{Q}h_z = \mathcal{D}_{\EuScript{C}^\perp} \xi_z + i \nu \cdot q_z \\
& \EuScript{Q} \tilde{q} = 0, \quad \EuScript{Q} \sigma = \mathcal{D}_{\EuScript{C}^\perp} \tilde{q}, \quad \EuScript{Q} h = \mathcal{D}_{\EuScript{C}^\perp} \sigma + i \nu \cdot \tilde{q}, \\
& \EuScript{Q} \chi = h^\dagger , \quad \EuScript{Q} h^\dagger = 0,\quad \EuScript{Q}  \eta_{\bar{z}} = h_{\bar{z}} ,\quad \EuScript{Q} h_{\bar{z}} =0, \\
&\EuScript{Q} \tilde{q}^\dagger =\gamma, \quad \EuScript{Q} \gamma = 0, \quad \EuScript{Q} q_{\bar{z}} = \zeta_{\bar{z}} \quad \EuScript{Q} \zeta_{\bar{z}} =0 .
\end{split}
\end{align}
Also in terms of the re-defined fields, the hypermultiplet action can be written as a linear combination of $\EuScript{Q}$-closed part and a $\EuScript{Q}$-exact part,
\begin{align}
S_{\text{hyp}} = S_{\text{hyp,cl}} + S_{\text{hyp,ext}},
\end{align}
where the $\EuScript{Q}$-exact part is given by
\begin{align} \label{eq:hypext}
\begin{split}
S_{\text{hyp,ext}} &= \EuScript{Q}\, \left\{ \frac{1}{g^2} \int_{\EuScript{C}} d^2 z \int_{\EuScript{C}^\perp}   \eta_{\bar{z}} \left(\star_{\EuScript{C}^\perp}  h_z -\frac{i}{2} D_z \tilde{q}^\dagger \right) + \chi \left( \star_{\EuScript{C}^\perp} h +\frac{i}{2} D_z q_{\bar{z}} \right)  -\frac{1}{2} \left( \tilde{q}^\dagger \alpha \cdot \tilde{q} + q_{\bar{z}} \alpha \cdot q_z \right) \right. \\
&\quad\quad\quad\quad\quad\quad \left. - \bar{\mathcal{D}}_{\EuScript{C}^\perp} \tilde{q}^\dagger \wedge \star_{\EuScript{C}^\perp} \sigma - \bar{\mathcal{D}}_{\EuScript{C}^\perp} q_{\bar{z}} \wedge \star_{\EuScript{C}^\perp} \xi_z  -\frac{1}{2} q_{\bar{z}} \mu_z \cdot \tilde{q}^\dagger \right\} ,
\end{split}
\end{align}
whereas the $\EuScript{Q}$-closed part is given by
\begin{align} \label{eq:hypcl}
\begin{split}
S_{\text{hyp,cl}} = \frac{8i}{g^2} \int_{\EuScript{C}} d^2 z \int_{\EuScript{C}^\perp}  \xi_z \wedge D_{\bar{z}} \sigma + h_z D_{\bar{z}} \tilde{q} - h D_{\bar{z}} q_z -i  q_z \mathsf{D}_{\bar{z}} \cdot \tilde{q} -i \tilde{q} \xi_z \wedge \rho_{\bar{z}} -i q_z \rho_{\bar{z}} \wedge \sigma.
\end{split}
\end{align}
Combining the vector multiplet action \eqref{eq:actvec} and the hypermultiplet action \eqref{eq:hypext}, \eqref{eq:hypcl}, we obtain the full action of the holomorphic-topological theory
\begin{align}
S= S_{\text{vec}} + S_{\text{hyp,cl}} + S_{\text{hyp,ext}}.
\end{align}
It should be reminded that, as firstly discovered in \cite{ref:kap}, the dependence on the metric on $\EuScript{C}^\perp$ and the K\"{a}hler form on $\EuScript{C}$ enters only through the $\EuScript{Q}$-exact terms, ensuring that the theory is topological along $\EuScript{C}^\perp$ and holomorphic along $\EuScript{C}$. Also note that we can absorb the gauge coupling in the $\EuScript{Q}$-closed part into the fields, so that the dependence on the gauge coupling also becomes absent. We also absorb the irrelevant numerical prefactors in some terms in $S_{\text{vec}}$ by rescaling the metric on $\EuScript{C}$. Now we may take the theory on a general product metric background of $\EuScript{C} \times \EuScript{C}^\perp$ while its component fields take values of appropriate differential forms.

For later use, it is convenient to define the following specific combinations of component fields:
\begin{align} \label{eq:cplx}
\begin{split}
&Q_z \equiv q_z + \xi_z + h_z \\
&\tilde{Q} = \tilde{q} + \sigma + h \\
& \EuScript{A}_{\bar{z}} = A_{\bar{z}}+\rho_{\bar{z}} + \mathsf{D}_{\bar{z}},
\end{split}
\end{align}
on which our localizing supercharge will act as the equivariant differential on $\EuScript{C}^\perp$. Note that we can re-write the $\EuScript{Q}$-closed part of the action \eqref{eq:hypcl} using these combinations as
\begin{align} \label{eq:actcl}
S_{\text{hyp,cl}} = 8i \int_{\EuScript{C}} d^2 z \int _{\EuScript{C}^\perp} Q_z \wedge (\partial_{\bar{z}} - i \EuScript{A}_{\bar{z}} \cdot ) \tilde{Q}.
\end{align}
where $\cdot$ denotes the action according to the representation under the gauge group. This expression will turn out to be useful in finding the action of the localized theory on $\EuScript{C}$.

\subsection{$\Omega$-deformation}
Suppose there is a vector field $V = \text{Vect}(\EuScript{C}^\perp)$ which generates an isometry on $\EuScript{C}^\perp$. The $\Omega$-deformation can be defined at the level of supersymmetry variations of component fields, so that the deformed supercharge squares to this isometry plus possibly a gauge transformation, in a similar manner with \cite{ref:nek, ref:cosyagi} for two-dimensional theories. For the case at hand, the holomorphic-topological theory on $\EuScript{C} \times \EuScript{C}^\perp$, we can deform the supersymmetry variations in \eqref{eq:susyvecundef} and \eqref{eq:susyhypundef} as
\begin{align} \label{eq:susyvecdef}
\begin{split}
&\EuScript{Q}_\varepsilon \mathcal{A} = \varepsilon \iota_V \nu, \quad \EuScript{Q}_\varepsilon \bar{\mathcal{A}} = \lambda - \varepsilon \iota_V \nu, \\
& \EuScript{Q}_\varepsilon \lambda = 2\varepsilon \iota_V F - 2 i \varepsilon D_{\EuScript{C}^\perp} \iota_V \phi, \quad \EuScript{Q}_\varepsilon \nu = \mathcal{F}, \\
& \EuScript{Q}_\varepsilon \alpha = \mathsf{D} , \quad \EuScript{Q}_\varepsilon \mathsf{D} = \varepsilon \iota_V \mathcal{D}_{\EuScript{C}^\perp} \alpha, \\
& \EuScript{Q}_\varepsilon A_{\bar{z}} = \varepsilon \iota_V \rho_{\bar{z}} ,\quad \EuScript{Q}_\varepsilon A_z = \theta_z, \\
& \EuScript{Q}_\varepsilon \rho_{\bar{z}} = \mathcal{F}_{\bar{z}} + \varepsilon \iota_V \mathsf{D}_{\bar{z}} , \quad \EuScript{Q}_\varepsilon \theta_z = \varepsilon \iota_V \mathcal{F}_z, \\
& \EuScript{Q}_\varepsilon \mathsf{D}_{\bar{z}} = \mathcal{D}_{\EuScript{C}^\perp} \rho_{\bar{z}} +D_{\bar{z}} \nu, \quad \EuScript{Q}_\varepsilon \mu_z = \mathsf{D}_z, \quad \EuScript{Q}_\varepsilon \mathsf{D}_z = \varepsilon \mathcal{D}_{\EuScript{C}^\perp} \iota_V \mu_z,
\end{split}
\end{align}
for the vector multiplet and
\begin{align} \label{eq:susyhypdef}
\begin{split}
& \EuScript{Q}_{\varepsilon} q_z =\varepsilon \iota_V \xi_z , \quad \EuScript{Q}_{\varepsilon} \xi_z =\mathcal{D}_{\EuScript{C}^\perp} q_z + \varepsilon \iota_V h_z , \quad \EuScript{Q}_{\varepsilon}h_z = \mathcal{D}_{\EuScript{C}^\perp} \xi_z + i\nu \cdot q_z  \\
& \EuScript{Q}_{\varepsilon} \tilde{q} = \varepsilon \iota_V \sigma, \quad \EuScript{Q}_{\varepsilon} \sigma = \mathcal{D}_{\EuScript{C}^\perp} \tilde{q} + \varepsilon\iota_V h, \quad \EuScript{Q}_{\varepsilon} h = \mathcal{D}_{\EuScript{C}^\perp} \sigma+i \nu \cdot \tilde{q}, \\
& \EuScript{Q}_{\varepsilon} \chi  = h ^\dagger , \quad \EuScript{Q}_{\varepsilon} h ^\dagger = \varepsilon \mathcal{D}_{\EuScript{C}^\perp} \iota_V \chi ,\quad  \EuScript{Q}_{\varepsilon}  \eta_{\bar{z}} = h_{\bar{z}} ,\quad \EuScript{Q}_{\varepsilon} h_{\bar{z}} = \varepsilon \mathcal{D}_{\EuScript{C}^\perp} \iota_V  \eta_{\bar{z}}, \\
&\EuScript{Q}_{\varepsilon} \tilde{q}^\dagger =\gamma, \quad \EuScript{Q}_{\varepsilon} \gamma = \varepsilon \iota_V \mathcal{D}_{\EuScript{C}^\perp} \tilde{q}^\dagger, \quad \EuScript{Q}_\varepsilon q_{\bar{z}} = \zeta_{\bar{z}} \quad \EuScript{Q}_\varepsilon \zeta_{\bar{z}} =\varepsilon \iota_V \mathcal{D}_{\EuScript{C}^\perp} q_{\bar{z}} .
\end{split}
\end{align}
for each hypermultiplet. Note that
\begin{align} \label{eq:omegasq}
\EuScript{Q}_\varepsilon^2 = \varepsilon(\mathcal{D}_{\EuScript{C}^\perp} \iota_V + \iota_V \mathcal{D}_{\EuScript{C}^\perp}) = \varepsilon \mathcal{L}_V + \text{Gauge} [\varepsilon \iota_V \mathcal{A}],
\end{align} 
where the first term is the Lie derivative with respect to the vector field $V$ and the second term is the infinitesimal gauge transformation generated by $\varepsilon \iota_V \mathcal{A}$. Hence the deformed supercharge squares to an isometry generated by $V$ plus a gauge transformation. Also it is immediate that $\EuScript{Q}_\varepsilon$ reduces to the original holomorphic-topological supercharge $\EuScript{Q}$ when $\varepsilon =0$. Hence $\EuScript{Q}_\varepsilon$ indeed implements the $\Omega$-deformation of the holomorphic-topological theory on $\EuScript{C}\times \EuScript{C}^\perp$ with respect to the isometry $V$.

We should correspondingly deform the action so that it is annihilated by the deformed supercharge. The action for the vector multiplet can be taken as the variation under the deformed supercharge of the same expression:
\begin{align} \label{eq:exact1}
\begin{split}
S_{\text{vec}, \varepsilon}&= \EuScript{Q}_\varepsilon \, \int_{\EuScript{C}} d^2 z \int_{\EuScript{C}^\perp} \text{Tr}\, \left[ -   \bar{\mathcal{F}} \star_{\EuScript{C}^\perp} \nu - \alpha \left( \star_{\EuScript{C}^\perp} \mathsf{D} -2i  D_{\EuScript{C}^\perp} \star_{\EuScript{C}^\perp} \phi - \star_{\EuScript{C}^\perp} F_{z\bar{z}} \right) \right. \\
&  \quad\quad\quad\quad\quad\quad\quad\quad\quad\quad \left. +  \bar{\mathcal{F}}_z  \wedge \star_{\EuScript{C}^\perp} \rho_{\bar{z}} +  \mu_z \star_{\EuScript{C}^\perp} \mathsf{D}_{\bar{z}} \right].
\end{split}
\end{align}
Similarly the $\EuScript{Q}$-exact part of the hypermultiplet action can be modified to:
\begin{align} \label{eq:exact2}
\begin{split}
S_{\text{hyp,ext},\varepsilon} &= \EuScript{Q}_\varepsilon\, \int_{\EuScript{C}} d^2 z \int_{\EuScript{C}^\perp}   \eta_{\bar{z}} \left(\star_{\EuScript{C}^\perp}  h_z -\frac{i}{2} D_z \tilde{q}^\dagger \right) + \chi \left( \star_{\EuScript{C}^\perp} h +\frac{i}{2} D_z q_{\bar{z}} \right)  -\frac{1}{2} \left( \tilde{q} \alpha \cdot \tilde{q}^\dagger + q_{\bar{z}} \alpha \cdot q_z \right) \\
&\quad\quad\quad\quad\quad\quad - \bar{\mathcal{D}}_{\EuScript{C}^\perp} \tilde{q}^\dagger  \wedge \star_{\EuScript{C}^\perp} \sigma -  \bar{\mathcal{D}}_{\EuScript{C}^\perp} q_{\bar{z}}   \wedge \star_{\EuScript{C}^\perp} \xi_z  -\frac{1}{2} q_{\bar{z}} \mu_z \cdot \tilde{q}^\dagger .
\end{split}
\end{align}
Since $V$ generates an isometry on $\EuScript{C}^\perp$, $\mathcal{L}_V$ leaves the metric invariant and commutes with $\star_{\EuScript{C}^\perp}$. Hence \eqref{eq:omegasq} guarantees that these actions are $\EuScript{Q}_\varepsilon$-invariant.

Finally, it is not difficult to check that the $\EuScript{Q}_\varepsilon$-variation of the $\EuScript{Q}_\varepsilon$-closed part of the action \eqref{eq:actcl} is now non-zero but a total derivative on $\EuScript{C}^\perp$. Thus it can be written as a contribution from the boundary $\partial \EuScript{C}^\perp$. We can simply add a boundary term to the action to cancel this \cite{ref:nek}, yielding
\begin{align} \label{eq:actcl2}
S_{\text{hyp,cl},\varepsilon} = 8i \left( \int_{\EuScript{C}} d^2 z \int _{\EuScript{C}^\perp} Q_z \wedge (\partial_{\bar{z}} - i \EuScript{A}_{\bar{z}} \cdot ) \tilde{Q} + \frac{1}{\varepsilon} \int_{\EuScript{C}} d^2 z \int_{\partial \EuScript{C}^\perp} V^{\vee} \; q_z D_{\bar{z}} \tilde{q} \right),
\end{align}
where $V^\vee$ is the one-form satisfying $\iota_V V^\vee=1$ and $\mathcal{L}_V V^\vee =0$ on $\EuScript{C}^\perp$. Hence the $\EuScript{Q}_\varepsilon$-invariance of the action is now established.

\subsection{Gauge-fixing}
Gauge-fixing is needed to properly evaluate the path integral. We implement the gauge-fixing by the standard BRST procedure. We introduce a ghost $c$, an antighost $\bar{c}$, and an auxiliary field $p$ which are in adjoint representation of the gauge group. The BRST transformations of these fields are
\begin{align}
\begin{split}
&\EuScript{Q}_B  c = -\frac{i}{2} \{ c,c \} , \quad \EuScript{Q}_B  \bar{c} = p, \quad \EuScript{Q}_B  p =0, \\
&\EuScript{Q}_B X = \text{Gauge}[c] X,
\end{split}
\end{align} 
where $X$ denotes all other component fields introduced in previous section. We also postulate the $\EuScript{Q}_\varepsilon$-variations for these fields as
\begin{align}
\EuScript{Q}_\varepsilon c = - \varepsilon \iota_V \mathcal{A},\quad \EuScript{Q}_\varepsilon \bar{c} = 0, \quad \EuScript{Q}_\varepsilon p = \varepsilon \iota_V d_{\EuScript{C}^\perp} \bar{c}.
\end{align}
Now we define a new supercharge $\hat{\EuScript{Q}}$ as the combination of the $\Omega$-deformed supercharge and the BRST supercharge, $\hat{\EuScript{Q}} = \EuScript{Q}_\varepsilon + \EuScript{Q}_B$. Then we observe that
\begin{align}
\hat{\EuScript{Q}}^2 = \varepsilon( d_{\EuScript{C}^\perp} \iota_V + \iota_V d_{\EuScript{C}^\perp}) = \varepsilon \mathcal{L}_V
\end{align}
for all fields. Note that the supercharge now squares to the isometry generated by $V$ without any gauge transformation. We use this supercharge $\hat{\EuScript{Q}}$ to construct our cohomological field theory.

Since \eqref{eq:exact1} and \eqref{eq:exact2} are defined as $\EuScript{Q}_\varepsilon$-variations of gauge invariant expressions, they are also automatically $\hat{\EuScript{Q}}$-exact:
\begin{align} 
\begin{split}
S_{\text{vec}, \varepsilon}&= \hat{\EuScript{Q}} \, \int_{\EuScript{C}} d^2 z \int_{\EuScript{C}^\perp} \text{Tr}\, \left[ -   \bar{\mathcal{F}} \star_{\EuScript{C}^\perp} \nu - \alpha \left( \star_{\EuScript{C}^\perp} \mathsf{D} -2i  D_{\EuScript{C}^\perp} \star_{\EuScript{C}^\perp} \phi - \star_{\EuScript{C}^\perp} F_{z\bar{z}} \right) \right. \\
&  \quad\quad\quad\quad\quad\quad\quad\quad\quad\quad \left. + \bar{\mathcal{F}}_z   \wedge \star_{\EuScript{C}^\perp} \rho_{\bar{z}} + \mu_z \star_{\EuScript{C}^\perp} \mathsf{D}_{\bar{z}} \right] \\
S_{\text{hyp,ext},\varepsilon} &= \hat{\EuScript{Q}}\, \int_{\EuScript{C}} d^2 z \int_{\EuScript{C}^\perp}   \eta_{\bar{z}} \left(\star_{\EuScript{C}^\perp}  h_z -\frac{i}{2} D_z \tilde{q}^\dagger \right) + \chi \left( \star_{\EuScript{C}^\perp} h +\frac{i}{2} D_z q_{\bar{z}} \right)  -\frac{1}{2} \left( \tilde{q} \alpha \cdot \tilde{q}^\dagger + q_{\bar{z}} \alpha \cdot q_z \right) \\
&\quad\quad\quad\quad\quad\quad - \bar{\mathcal{D}}_{\EuScript{C}^\perp} \tilde{q}^\dagger  \wedge \star_{\EuScript{C}^\perp} \sigma -  \bar{\mathcal{D}}_{\EuScript{C}^\perp} q_{\bar{z}}  \wedge \star_{\EuScript{C}^\perp} \xi_z  -\frac{1}{2} q_{\bar{z}} \mu_z \cdot \tilde{q}^\dagger ,
\end{split}
\end{align}
It is also clear that \eqref{eq:actcl2} is $\hat{\EuScript{Q}}$-closed since it is gauge-invariant. To gauge-fix we introduce another $\hat{\EuScript{Q}}$-exact term to the action
\begin{align}
S_{\text{fix}} = \hat{\EuScript{Q}} \int \text{Tr}\, \bar{c} \, G_{\text{fix}},
\end{align}
where $G_{\text{fix}}$ is a properly chosen gauge-fixing function. We will take the standard Lorentz gauge-fixing function
\begin{align}
G_{\text{fix}} = \nabla_{\mu} A^\mu, \quad \mu = w, \bar{w},
\end{align} 
where $\nabla$ the Levi-Civita connection on $\EuScript{C}^\perp$, while keeping the gauge redundancy on $\EuScript{C}$ intact. We will fix the residual gauge redundancy after localizing the theory onto $\EuScript{C}$.\footnote{The zero mode of the ghost $c$ would have been constant, but we absorb it to the ghost for the gauge fixing on $\EuScript{C}$ which will be introduced later. Hence we take the zero mode of $c$ here to be zero.}

\subsection{Localization}
For the purpose of recovering the chiral CFT on $\EuScript{C}$, we will take $\EuScript{C}^\perp = \mathbb{R}^2$ from now on. Let us analyze the localization locus of the path integral. The auxiliary fields $\mathsf{D}_z$, $h_{\bar{z}}$, and $h^\dagger$ only enters in the action in linear terms. Hence we can integrate them out to find
\begin{align} \label{eq:aux}
\begin{split}
&\mathsf{D}_{\bar{z}} = \frac{1}{2} \star_{\EuScript{C}^\perp} q_{\bar{z}} \tilde{q}^\dagger \\
&h_{z} = \frac{i}{2} \star_{\EuScript{C}^\perp} D_z \tilde{q}^\dagger \\
& h = -\frac{i}{2} \star_{\EuScript{C}^\perp} D_z q_{\bar{z}}.
\end{split} 
\end{align}
By completing the square for the terms involving $\mathsf{D}$, we also find
\begin{align} \label{eq:dterm}
\mathsf{D} = \frac{1}{2} \left( \star_{\EuScript{C}^\perp} D_{\EuScript{C}^\perp} \star_{\EuScript{C}^\perp} \phi + \frac{1}{2} \left(-i F_{z\bar{z}} +\tilde{q} \tilde{q}^\dagger +q_{\bar{z}} q_z  \right) \right).
\end{align}
The localization locus is given by the fixed point set of the supersymmetry variations. Hence we set the right hand sides of \eqref{eq:susyvecdef} and \eqref{eq:susyhypdef} to zero. Thus we have, among other equations,
\begin{align} \label{eq:fpeq}
\begin{split}
&\mathcal{F}=0, \quad \iota_V F -  i  D_{\EuScript{C}^\perp} \iota_V \phi= 0, \quad \mathsf{D}=0, \\ 
&\mathcal{F}_{\bar{z}} + \varepsilon \iota_V \mathsf{D}_{\bar{z}} =0, \quad \mathcal{D}_{\EuScript{C}^\perp} q_{z} + \varepsilon \iota_V h_z  =0, \quad \mathcal{D}_{\EuScript{C}^\perp} \tilde{q} + \varepsilon \iota_V h=0.
\end{split}
\end{align}
From the equations in the first row we get $F=0$, and since $\EuScript{C}^\perp =\mathbb{R}^2$ is simply-connected we can choose a gauge to set $A=0$. Applying this to the equations in the second row yields, among other equations, $D_{\bar{z}} \phi = \phi \cdot q_z =  \phi \cdot \tilde{q}=0$, implying the gauge transformation generated by $\phi$ is zero. By making a genericity assumption for $A_{\bar{z}}$, $q_z$, and $\tilde{q}$, we are led to $\phi =0 $. Then we arrive at 
\begin{align} \label{eq:gauge}
\mathcal{A} =0.
\end{align}
With \eqref{eq:aux} and \eqref{eq:gauge} the rest of the equations in the second row of \eqref{eq:fpeq} yield
\begin{align} \label{eq:flow1}
\begin{split}
&d_{\EuScript{C}^\perp} A_{\bar{z}} = - \frac{1}{2} \varepsilon \iota_V \star_{\EuScript{C}^\perp} q_{\bar{z}} \tilde{q}^\dagger \\
& d_{\EuScript{C}^\perp} q_{z} =  -\frac{i}{2} \varepsilon \iota_V \star_{\EuScript{C}^\perp} D_z \tilde{q}^\dagger \\
& d_{\EuScript{C}^\perp} \tilde{q} = \frac{i}{2} \varepsilon \iota_V \star_{\EuScript{C}^\perp} D_z q_{\bar{z}}.
\end{split}
\end{align}
Let us introduce the polar coordinate on $\EuScript{C}^\perp = \mathbb{R}^2$, where the flat metric on $\EuScript{C}^\perp$ is simply written as $ds_{\EuScript{C}^\perp} ^2 = dr^2 + r^2 d\varphi^2$. Then our generator of the isometry is $V = \partial_\varphi$. The equations \eqref{eq:flow1} can be written in the polar coordinates as
\begin{align}
\begin{split}
&\partial_\varphi A_{\bar{z}} =0, \quad \partial_\varphi q_z =0 , \quad \partial_\varphi \tilde{q} =0,\\
&\partial_r A_{\bar{z}} = -\frac{1}{2}\varepsilon r q_{\bar{z}} \tilde{q}^\dagger, \quad \partial_r q_z =- \frac{i}{2}\varepsilon r D_z \tilde{q}^\dagger, \quad \partial_r \tilde{q} =  \frac{i}{2} \varepsilon r D_z q_{\bar{z}}.
\end{split}
\end{align}
By re-defining the radial coordinate by $t= \varepsilon \bar\varepsilon \frac{r^2}{2}$, the equations in the second line become
\begin{align} \label{eq:gradflow}
\partial_t A_{\bar{z}} = -\frac{1}{ 2\bar{\varepsilon}} q_{\bar{z}} \tilde{q}^\dagger, \quad \partial_t  q_z = - \frac{i}{2\bar\varepsilon} D_z \tilde{q}^\dagger, \quad \partial_t \tilde{q} = \frac{i}{2\bar\varepsilon} D_z q_{\bar{z}}.
\end{align}
Solutions to these equations are precisely the gradient trajectories on which two-dimensional $B$-model on $\mathbb{R}^2$ localizes, as discussed in \cite{ref:nek} with its full detail, generated by the function $\text{Re}\,\left(\frac{\EuScript{W}}{\varepsilon}\right)$ where $\EuScript{W}$ is the holomorphic superpotential. For the case at hand, one could view the four-dimensional holomorphic-topological theory on $\EuScript{C}\times \EuScript{C}^\perp$ as a two-dimensional $B$-twisted gauge theory on $\EuScript{C}^\perp$, as done in \cite{ref:cosyagi} for the six-dimensional holomorphic-topological theory to obtain the four-dimensional Chern-Simons theory. The superpotential has to be chosen as $\EuScript{W} = \int_{\EuScript{C}} d^2 z \, q_z D_{\bar{z}} \tilde{q}$ to reproduce the four-dimensional holomorphic-topological theory in the $\varepsilon \to 0$ limit. As we approch to infinity $t \to \infty$, $A_{\bar{z}}$, $q_z$, and $\tilde{q}$ should end on the critical points $\{ d  \EuScript{W} =0 \}$ of the superpotential to guarantee that the action \eqref{eq:actcl} does not diverge \cite{ref:nek}.\footnote{Generally, when the critical points are non-isolated we can choose a Lagrangian submanifold of the critical points to have a constant one-loop determinant \cite{ref:cosyagi}, so we make such a choice here.}

We have to regard these flow equations as defined on the fields with complexified gauge group. Note that from \eqref{eq:dterm} and \eqref{eq:fpeq} we get
\begin{align}
\boldsymbol{\mu} \equiv -iF_{z\bar{z}} +\tilde{q} \tilde{q}^\dagger +q_{\bar{z}} q_z =0.
\end{align}
This equation is invariant under real gauge transformations, but not under the non-compact part of the complex gauge transformations. Also, generic complexified fields can be transformed into the fields lying in $\boldsymbol\mu^{-1} (0)$ by making a complex gauge transformation.\footnote{There is an issue of stability here. When the closure of the orbit of complexified gauge transformations intersects with $\boldsymbol\mu^{-1} (0)$, such a locus is called semistable. We are restricting to the semistable locus.} In other words, the restriction of the fields to the level set $\boldsymbol\mu^{-1} (0)$ can be viewed as gauge-fixing the non-compact part of the complexified gauge symmetry. In turn, we may just omit this equation and compensate it with complexifying the gauge group for the fields appearing in the flow equations \eqref{eq:gradflow}. This is precisely analoguous to the complexification of the gauge group for the analytically continued Chern-Simons theory \cite{ref:wit1, ref:wit2, ref:wit3}.

Now that we identified the localization locus, let us evaluate the effective action of the localized path integral. Recall that our theory is holomorphic along $\EuScript{C}$, so that the localized path integral should define a two-dimensional chiral CFT on $\EuScript{C}$. Also note that the localization locus does not contain any non-trivial topological sector of gauge field configurations, so that all we have to do is to evaluate the $\hat{\EuScript{Q}}$-closed part of the action on the localization locus properly. This can be accomplished by performing an equivariant integration on $\EuScript{C}^\perp = \mathbb{R}^2$ for the action integral \eqref{eq:actcl} as follows.

To facilitate the equivariant integration, it is crucial to note that $\hat{\EuScript{Q}}$ acts on the combinations \eqref{eq:cplx} as the equivariant differential $d_{\EuScript{C}^\perp} + \varepsilon \iota_V$ on $\EuScript{C}^\perp = \mathbb{R}^2$ plus a \textit{gauge covariant} contribution:
\begin{align} \label{eq:qhatcplx}
\begin{split}
&\hat{\EuScript{Q}} Q_z = (d_{\EuScript{C}^\perp} + \varepsilon \iota_V - i C \cdot) Q_z \\ 
&\hat{\EuScript{Q}} \tilde{Q} = (d_{\EuScript{C}^\perp} + \varepsilon \iota_V - i C \cdot) \tilde{Q} \\
&\hat{\EuScript{Q}} \EuScript{A}_{\bar{z}} = (d_{\EuScript{C}^\perp} +\varepsilon \iota_V - i C  \cdot) \EuScript{A}_{\bar{z}} -\partial_{\bar{z}} C,
\end{split}
\end{align}
where
\begin{align}
C \equiv c + \mathcal{A} + \nu
\end{align}
acts as if it is a \textit{gauge connection} for those complexes. Note that the last term in the third equality ensures that $\partial_{\bar{z}} - i \EuScript{A}_{\bar{z}}$ can be treated as a \textit{covariant derivative} as far as $\hat{\EuScript{Q}}$-variation is concerned, namely it preserves the gauge charge:
\begin{align}
\hat{\EuScript{Q}} \left( (\partial_{\bar{z}} - i \EuScript{A}_{\bar{z}} \cdot) \tilde{Q} \right ) =   (d_{\EuScript{C}^\perp} + \varepsilon \iota_V - i C \cdot ) \left((\partial_{\bar{z}} - i \EuScript{A}_{\bar{z}}\cdot) \tilde{Q} \right).
\end{align}
This would have failed without the last term of the third equality of \eqref{eq:qhatcplx}. Therefore, $\hat{\EuScript{Q}}$ acts as the equivariant differential on the gauge-invariant combination,
\begin{align}
\hat{\EuScript{Q}}\,\left(Q_z \wedge (\partial_{\bar{z}} - i \EuScript{A}_{\bar{z}} \cdot ) \tilde{Q}\right) = (d_{\EuScript{C}^\perp} +\varepsilon \iota_V) \left(Q_z \wedge (\partial_{\bar{z}} - i \EuScript{A}_{\bar{z}} \cdot ) \tilde{Q}\right).
\end{align}
In other words, $ Q_z \wedge (\partial_{\bar{z}} - i \EuScript{A}_{\bar{z}} \cdot) \tilde{Q}$ is equivariantly closed when it is viewed as an element in the $\hat{\EuScript{Q}}$-cohomology. Hence we apply the Atiyah-Bott equivariant localization formula for the bulk term in the action integral \eqref{eq:actcl2}. Since we have included the infinity $\partial \EuScript{C}^\perp = \{ t=\infty \}$ in our consideration, we have to regard $\EuScript{C}^\perp \cup \partial \EuScript{C}^\perp $ as the one-point compactification $S^2$ whose fixed points with respect to $V = \partial_\varphi$ are precisely the origin and the infinity, $t=0$ and $t=\infty$. The contribution from $t=\infty$ cancels the contribution from the second term in \eqref{eq:actcl2} ($V^\vee = d \varphi$), leaving only the contribution from the origin $t=0$ (while absorbing irrelevant numerical constant in front into $q_z$ and $\tilde{q}$):
\begin{align} \label{eq:locact}
S_{\text{hyp,cl},\varepsilon} = \frac{1}{\varepsilon} \int_{\EuScript{C}} d^2 z \, q_z D_{\bar{z}} \tilde{q},
\end{align}
where $q = q_z dz$ is a $(1,0)$-form in the representation $R$ and $\tilde{q}$ is a $0$-form in the representation $\bar{R}$ of the gauge group, respectively. As mentioned above, $q_z$, $\tilde{q}$, and $A_{\bar{z}}$ are understood here as solutions to the gradient trajectory equations \eqref{eq:gradflow} evaluated at the origin of $\EuScript{C}^\perp$, $w=\bar{w}=0$ (i.e. $t=0$). Therefore, the result of the localization is the two-dimensional path integral on $\EuScript{C}$ defined by the action \eqref{eq:locact}, and the integration cycle is the field configurations that can be reached by the gradient flow \eqref{eq:gradflow} emanating from the critical points. Note that this integration cycle ensures the convergence of the path integral even though the action is now complex-valued, and it is again precisely analoguous to how the convergence of the path integral is guaranteed for the analytically continued Chern-Simons theory with complexified gauge group \cite{ref:wit1, ref:wit2, ref:wit3}. Also, note that the $\Omega$-deformation parameter $\varepsilon$ appears in the denominator of the action since $\EuScript{C}^\perp = \mathbb{R}^2$ has the unit weight under the isometry of $V=\partial_\varphi$. Consequently $\varepsilon$ plays the role of the Planck constant of the localized theory on $\EuScript{C}$, which therefore appears in the numerator of the OPEs. Hence we confirm the identification of the non-commutative deformation parameter and the $\Omega$-deformation parameter.

Now we choose to fix the residual gauge by the gauge-fixing function $A_{\bar{z}} =0$, yielding the gauge fixing term in the action
\begin{align}
\frac{1}{\varepsilon} \int_{\EuScript{C}} d^2z \, \text{Tr}\, ( -p_z A_{\bar{z}} + b_z D_{\bar{z}} c ).
\end{align}
Hence when the auxiliary field $p_z$ is integrated out, we are left with
\begin{align}
\frac{1}{\varepsilon} \int_{\EuScript{C}} \left(\text{Tr}\,  b\bar{\partial} c + \sum_i q^i \bar{\partial} \tilde{q}^i \right),
\end{align}
where $i$ enumerates all the hypermultiplets that we coupled to the vector multiplet to make the original $\EuScript{N}=2$ theory superconformal. The algebra generated by the local operators of this theory are nothing but the chiral algebra of the standard $bc$-$\beta\gamma$ system with the BRST charge
\begin{align}
Q_{\text{BRST}} = \frac{1}{ \varepsilon} \oint \frac{dz}{2\pi i} \left( \text{Tr}\, bcc - \sum_i q^i c \tilde{q}^i \right).
\end{align}
Hence we arrive at the result expected from \cite{ref:bllprr}.

\section{Superconformal indices and vacuum characters} \label{sec:ind}
As a consequence of the SCFT/VOA correspondence in \cite{ref:bllprr}, the Schur index of the $\EuScript{N}=2$ SCFT and the vacuum character of the chiral algebra are identified by directly comparing their state-counting formulas. Here we discuss how the $\Omega$-deformation approach provides a path integral point of view on the identification.

\subsection{Schur index of $\EuScript{N}=2$ SCFT}
The Schur index is defined by the Schur limit of the $\EuScript{N}=2$ superconformal index \cite{ref:grry}. It is given as
\begin{align}
\mathcal{I}_S = \text{Tr}_{\mathcal{H}_S} \, (-1)^F q^{E-R},
\end{align}
where $E$ is the scaling dimension and $R$ is the Cartan of the $SU(2)_R$ R-symmetry as before. The trace is over the $\frac{1}{4}$-BPS states satisfying
\begin{align}
\begin{split}
\mathcal{H}_S: \quad &E-(j_1+j_2)-2R =0, \quad j_1 - j_2 +r =0.
\end{split}
\end{align}
The operators corresponding to these states are called Schur operators. It is straightforward to compute the single-letter indices for the vector multiplet and the hypermultiplet by finding those operators in the component fields. The full index is simply given by the plethystic exponential of the sum of all the single-letter indices, integrated over the gauge group. We will not reproduce the exact forms of those expressions here.

\subsection{Schur index and vacuum character}
In \cite{ref:pp, ref:df}, the Schur index was derived by supersymmetric localization of $\EuScript{N}=2$ SCFT partition function on $S^3 \times S^1$, up to a multiplicative factor of the Casimir energy. We start from the following metric background:
\begin{align}
ds^2 = l^2  \cos^2 \theta\left( d\psi - (\beta_1 +\beta_2) dt \right)^2 +l^2 \sin^2 \theta \left( d\varphi -(\beta_1-\beta_2) dy \right)^2 +l^2 d\theta^2 -l^2( \tau+(\beta_1 +\beta_2))^2 dy^2,
\end{align}
where $\psi$, $\varphi$, and $y$ are periodic coordinates with period $2\pi$ and $\theta \in [0, \frac{\pi}{2}]$. $l$ is the radius of the three-sphere which was written as a torus fibration over the $\theta$-interval. It was shown in \cite{ref:pp} that the variations of $\beta_1$ and $\beta_2$ do not affect the partition function. Then $\beta_1$ and $\beta_2$ were chosen to be real and $\text{Re}\, \tau = -(\beta_1+\beta_2)$ so that above metric restricts to the K\"{a}hler metric on the torus at $\theta=0$,
\begin{align}
ds_{\EuScript{C}} ^2 = l^2 dz d\bar{z},
\end{align}
where $z=\psi + \tau y$ and $\bar{z} = \psi - \bar{\tau} y$. To find the supercharge preserved under the non-trivial background, we need to solve the generalized Killing spinor equation. It was shown that there are four solutions to the generalized Killing spinor equation which generate an $\mathfrak{su} (1\vert1) \oplus \mathfrak{su} (1\vert1) $ superalgebra. Then we can choose a localizing supercharge $\mathcal{Q}$ in this superalgebra satisfying
\begin{align} \label{eq:charge}
\mathcal{Q} ^2 = \frac{1}{l} \mathcal{L}_{ \partial_\varphi} + R_r \left( -\frac{1}{l} \right) + \text{Gauge transformation},
\end{align}
where $R_r$ is the $U(1)_r$ rotation. The path integral localizes onto the torus $\theta=0$ as a result of the localization with respect to $\mathcal{Q}$, yielding the expression of the Schur indices as torus partition functions of two-dimensional CFT, or the characters of vertex operator algebra.

It was discussed further in \cite{ref:df} that we can take a decompactification limit of the above setting, which zooms in the region around $\theta=0$, to find a direct connection to the $\Omega$-deformation picture which we have described throughout the present work. Let us set $\beta_1=\beta_2 =0$ for convenience. We re-define the coordinates as
\begin{align}
\tilde{y} = ly , \quad \tilde{\psi} = l\psi, \quad r= l \theta,
\end{align}
and take the $l \equiv \frac{1}{\varepsilon} \to \infty$ limit. Then the metric becomes
\begin{align}
ds^2 = \vert d\tilde\psi + \tau d\tilde{y} \vert^2  + dr^2 + r^2 d\varphi^2 .
\end{align}
This is precisely the product metric $\EuScript{C} \times \EuScript{C}^\perp$ with $\EuScript{C}$ now being another $ \mathbb{R}^2 =(\tilde{\psi}, \tilde{y})$. Moreoever, we recognize the supercharge $\mathcal{Q}$ used here implements the $\Omega$-deformation on $\EuScript{C}^\perp$, in the sense of the relation \eqref{eq:charge}.

Hence we can make a direct connection between the partition functions in four-dimension and two-dimension. First we localize the $\Omega$-deformed holomorphic-topological theory along $\EuScript{C}^\perp$, leaving the chiral CFT of the gauged symplectic bosons on the torus $\EuScript{C}$. The above argument shows that we actually recover the $S^3 \times S^1$ partition function of the four-dimensional theory, and thus leading to the identification of the $S^3 \times S^1$ partition function of the four-dimensinoal SCFT with the torus partition function of the two-dimensional chiral CFT. 

As just mentioned, in the localizing supercharge $\mathcal{Q}$, the $S^3 \times S^1 $ partition function of four-dimensinoal SCFT computes the Schur index \cite{ref:pp, ref:df} (in \cite{ref:df}, the multiplicative factor of Casimir energy is also matched). Also the torus partition functions of two-dimensional chiral CFT compute the characters of the chiral algebra. Hence we re-discover one of the consequences of the SCFT/VOA correspondence found in \cite{ref:bllprr}, the identification of the Schur index and the vacuum character, at the level of their path integral representations. For further discussions on the path integral representations of the Schur index and its identification with the VOA character, see \cite{ref:pp, ref:df}.

\section{Discussion} \label{sec:dis}

In the ($\mathcal{Q}+\mathcal{S}$)-cohomology construction of the chiral algebra \cite{ref:bllprr}, the $bc$-system is obtained by the cohomology of the Schur operators in the vector multiplets: the gaugini. In our notation, the relevant gaugini fields are precisely $\mu_z$ and $\theta_z$ in \eqref{eq:redefvec}. However, it is still not very clear how these fields are related to the $bc$-system in our construction which arises in the two-dimensional gauge fixing. The main problem is that neither $\mu_z$ nor $\theta_z$ is $\EuScript{Q}$-closed, so that it is not immediate to see how they come into play in the $\EuScript{Q}$-cohomological field theory. It would be nice if we can understand this issue more clearly.

An interesting observation was made in \cite{ref:song} for the SCFT/VOA correspondence at the level of $\EuScript{N}=2$ superconformal indices. It discovered a relation between the Macdonald index, a refinement of the Schur index, of $\EuScript{N}=2$ SCFTs and the refined character of VOAs. A conjectural construction of a filtration of the vacuum module was suggested, from which the refined character was defined by its associated graded vector space. In \cite{ref:br}, the construction of such filtrations was analyzed in great detail. It would be nice if we can understand this relation through the $\Omega$-deformation formulation of the chiral algebra. A path integral representation of the Macdonald index or the suggested refined character would be helpful for this study.

Finally, the $\Omega$-deformation approach to the chiral algebra discussed so far only applies to Lagrangian SCFTs. It was observed that in some cases there are $\EuScript{N}=1$ preserving deformations of $\EuScript{N}=2$ SQCDs such that the renormalization group flows from the deformed SQCDs to non-Lagrangian $\EuScript{N}=2$ SCFTs such as Argyres-Douglas theories and Minahan-Nemeschansky theories \cite{ref:marsong1, ref:marsong, ref:ams, ref:ass, ref:ams2}. It would be nice if we could find a way to apply the $\Omega$-deformation procedure to obtain the VOAs for non-Lagrangian SCFTs \cite{ref:xy}, perhaps by using such deformations.

\acknowledgments
SJ is grateful to Nikita Nekrasov for numerous discussions and supports, and to Christopher Beem and Leonardo Rastelli for collaboration in the initial stage of the work. SJ would like to thank Jean-Emile Bourgine, Tudor Dimofte, Sergei Gukov, Andrew Neitzke, and Savdeep Sethi for sharing helpful discussions and providing supports during his visit to Korea Institute for Advanced Study, University of California at Davis, California Institute of Technology, University of Texas at Austin, and University of Chicago. The work was supported in part by the NSF grant PHY 1404446 and also by the generous support of the Simons Center for Geometry and Physics. 

\appendix
\section{Conventions}
The spinor indices in $\psi_\alpha$ and $\tilde{\psi}^{\dot{\alpha}}$ are raised and lowered by
\begin{align}
\begin{split}
&\psi^\alpha = \epsilon^{\alpha\beta} \psi_\beta, \quad \tilde{\psi}_{\dot{\alpha}} = \epsilon_{\dot{\alpha} \dot{\beta}} \tilde{\psi}^{\dot{\beta}} \\
&\psi_\beta = -\psi^\alpha \epsilon_{\alpha\beta}, \quad \tilde\psi^{\dot\beta} = - \tilde\psi_{\dot\alpha} \epsilon^{\dot\alpha \dot\beta},
\end{split}
\end{align}
where $\epsilon^{12} = -\epsilon_{12} = \epsilon^{\dot{1} \dot{2}} =-\epsilon_{\dot{1}\dot{2}} =1$. We use the convention for the spinor index contraction
\begin{align}
\psi \chi = \psi^{\alpha} \chi_\alpha, \quad \tilde{\psi}\tilde{\chi} = \tilde{\psi}_{\dot{\alpha}} \tilde{\chi}^{\dot{\alpha}}.
\end{align}
The symplectic-Majorana spinors $\psi_A$ and $\tilde{\psi}_A$ are defined by
\begin{align}
(\psi_{\alpha A} )^\dagger = \epsilon^{AB}\epsilon^{\alpha\beta} \psi_{\beta B}, \quad (\tilde{\psi}_{\dot{\alpha}A})^\dagger =\epsilon^{AB} \epsilon ^{\dot{\alpha} \dot{\beta}} \tilde{\psi}_{\dot{\beta}B},
\end{align}
where the $SU(2)_R$ indices are raised and lowered as $X^A = \epsilon^{AB} X_B$ and $X_A = \epsilon_{AB} X^B$ with $\epsilon^{12}=-{\epsilon}_{12}=1$.

The $\sigma$-matrices are defined by
\begin{align}
\sigma^a _{\alpha \dot{\alpha}} = (i \vec{\tau} , \mathds{1})_{\alpha \dot{\alpha}}, \quad \tilde{\sigma}^{a \dot{\alpha}\alpha} = (-i\vec{\tau} , \mathds{1})^{\dot{\alpha}\alpha},
\end{align}
where $\vec{\tau}$ are the Pauli matrices.

\end{document}